\documentclass[%
 reprint,
 amsmath,amssymb,
 aps,
pra,
]{revtex4-2}

\usepackage{graphicx}
\usepackage{dcolumn}
\usepackage{bm}
\usepackage{amsmath}
\usepackage{bbm}
\usepackage{blkarray}
\usepackage{booktabs}
\usepackage{mathtools}
\begin{document}

\preprint{APS/123-QED}

\title{Symmetry constraints for vector scattering and transfer matrices containing evanescent components: energy conservation, reciprocity and time reversal}

\author{Niall Byrnes}
\author{Matthew R Foreman}%
 \email{Corresponding author: matthew.foreman@imperial.ac.uk}
\affiliation{%
 Blackett Laboratory, Department of Physics, Imperial College London, Prince Consort Road, London, SW7 2AZ, United Kingdom
}%

\date{\today}

\begin{abstract}
In this work we study the scattering and transfer matrices for electric fields defined with respect to an angular spectrum of plane waves. For these matrices, we derive the constraints that are enforced by conservation of energy, reciprocity and time reversal symmetry. Notably, we examine the general case of vector fields in three dimensions and allow for evanescent field components. Moreover, we consider fields described by both continuous and discrete angular spectra, the latter being more relevant to practical applications, such as optical scattering experiments. We compare our results to better-known constraints, such as the unitarity of the scattering matrix for far-field modes, and show that previous results follow from our framework as special cases. Finally, we demonstrate our results numerically with a simple example of wave propagation at a planar glass-air interface, including the effects of total internal reflection. Our formalism makes minimal assumptions about the nature of the scattering medium and is thus applicable to a wide range of scattering problems.

\end{abstract}

\maketitle


\section{Introduction}
Scattering and transfer matrices are important mathematical tools that have been applied in a variety of fields, including the transport of electrons in wires \cite{PhysRevB.22.3519, PhysRevB.23.6851}, telecommunications \cite{tulino2004random}, acoustics \cite{ PhysRevLett.102.084301, TO1979207} and photonic crystals \cite{Mingaleev:03, BELL1995306}. Within both formalisms, a scattering object or medium is viewed as a `black box' and the scattering and transfer matrices describe the coupling between different modes in the surrounding regions. 

In optics, the scattering matrix, which describes the reflection and transmission of electromagnetic modes by a system, is particularly powerful in the study of disordered media for which the underlying scattering potential is practically unknowable, but for which the reflected and transmitted fields are experimentally measurable \cite{RevModPhys.89.015005}. In many modern scattering experiments, the degrees of freedom of an electromagnetic field are explored through wavefront shaping using spatial light modulators, which has facilitated experimental measurements of scattering matrices of complex media for both scalar and vectorial light \cite{PhysRevLett.104.100601, YU201533, Tripathi:12}. Knowledge of these matrices and their statistical properties has revealed the existence of highly transmitting `open eigenchannels', even for optically thick systems \cite{PhysRevLett.101.120601, PhysRevB.83.134207, Kim2012}, and has greatly enhanced the prospect of imaging through multiple scattering media, such as biological tissue, through careful wavefront control and the exploitation of scattered field correlations, such as the memory effect \cite{Popoff2010, Choi:11, Kang2017, Bertolotti2012, Katz2014, memoryTrans, hsu}. Scattering matrices have also been employed in theoretical studies of random laser modes \cite{Andreasen:11}, information transfer through random media \cite{Byrnes_2020}, coherent backscattering \cite{Sheikhan_2012} and Anderson localization \cite{anderson}.

Alternatively, the transfer matrix describes coupling between modes on either side of a scattering medium, provided that the geometry of the system permits a meaningful identification of two opposite sides. Examples of such systems include optical waveguides \cite{Torner, Carpenter:16, 62888} and stratified media consisting of a series of contiguous slabs \cite{yeh2004optical, doi:10.1119/1.1308266}, such as superlattices \cite{PhysRevB.67.085318, PhysRevLett.80.2677} and multilayer thin-films \cite{Katsidis:02, azzam1977ellipsometry}. The primary advantage of transfer matrices over scattering matrices for layered systems is that a system's overall transfer matrix can be easily computed by taking the correctly-ordered matrix product of the transfer matrices of each separate layer \cite{1991RSPSA.435..185B}. In contrast, the corresponding composition law for scattering matrices is more computationally demanding \cite{Ko} and, in some instances, requires unphysical assumptions, such as neglecting multiple reflections between surfaces within a stratified medium \cite{Gu:93}. This property makes transfer matrices particularly well suited to numerical simulations \cite{LIU2019185, doi:10.1121/1.408152, privman1990finite} and theoretical studies, such as in photonic band structures \cite{doi:10.1080/09500349414550281} and mesoscopic scattering \cite{MELLO1988290, PhysRevB.44.3559}.

Traditionally, scattering and transfer matrices are defined only for modes that propagate to the far field. The exact values of the elements of these matrices depend strongly on the type of system being considered and may vary significantly from one example to another. Nevertheless, under rather general conditions, both matrices can be shown to obey certain mathematical constraints valid for large classes of scattering media. For example, it has long been known that a system that conserves energy, i.e. does not absorb or generate light, regardless of its microscopic configuration, must possess a scattering matrix that is unitary \cite{RevModPhys.69.731}. More recently, techniques such as scanning near-field optical microscopy have enabled studies in which the scattering of evanescent fields play a critical role, such as in single molecule near-field imaging \cite{Betzig1422}, scattering from plasmonic nanoantennas \cite{doi:10.1002/lpor.201500031}, particle tracking \cite{9250534} and near-field speckle imaging \cite{doi:10.1063/1.4976747}. Such studies have motivated the introduction of an extended version of the scattering matrix that is capable of describing scattering to and from evanescent field components. In a notable paper by Carminati et al., the mathematical constraints obeyed by the extended scattering matrix were explored for scalar waves under the conditions of conservation of energy, reciprocity and time reversal symmetry \cite{PhysRevA.62.012712}. The corresponding scattering matrix constraints due to reciprocity for vector evanescent waves has also been considered separately \cite{Carminati:98}. 

Matrix constraints such as those mentioned place limits on the set of all physically possible scattering and transfer matrices, and hence can serve as useful guides in determining whether a given experimental or simulated matrix satisfies the corresponding physical law. In addition, these constraints may also be useful in designing matrix-based models and simulations for scattering in complex media. Furthermore, random matrix theory, in which the set of constraints satisfied by a matrix is generally the only assumption made, has proven to be very fruitful at uncovering universal properties of random scattering media \cite{Byrnes_2020, mehta2004random, RevModPhys.69.731}. We therefore believe that an accurate knowledge of the constraints satisfied by both the scattering and transfer matrices 
is important for future studies of scattering problems.

Compared to the scattering matrix, theoretical analysis of the transfer matrix seems to have received less attention in the optics literature. Moreover, while previous works have explored matrix constraints pertinent to a continuous decomposition of an electric field containing an infinite set of modes, such as in a continuous angular spectrum decomposition \cite{Nieto-Vesperinas:86, vesperinas2006scattering}, the corresponding constraints satisfied by the scattering and transfer matrices defined with respect to a finite set of modes are equally important, particularly for experiments and simulations in which only a finite description is physically possible. The purpose of this paper is therefore to present a self-contained derivation of the set of constraints imposed upon the scattering and transfer matrices by conservation of energy, reciprocity and time reversal symmetry. Notably, our treatment takes full account of the vector nature of light and allows for evanescent components. A treatment of vector fields is essential in optics, as scattering typically gives rise to polarization mixing and depolarization, neither of which can be described within a scalar wave formalism. We describe arbitrary fields using a vector angular spectrum decomposition, first over a continuous range encompassing all possible wavevectors and then for a discrete angular spectrum containing a finite set of modes. The angular spectrum decomposition is particularly useful as it is able to conveniently discriminate between propagating and evanescent modes. This work builds upon previous results and, for the sake of literary continuity, we adopt similar notation and follow a similar format to that of Ref. \cite{PhysRevA.62.012712}. 

In Section II, we define the scattering and transfer matrices for a continuous angular spectrum of an electric field and derive the constraints imposed by the aforementioned conditions. In Section III, we introduce the scattering and transfer matrices for a discrete angular spectrum and derive the associated constraints from the continuous case. We then compare our results to previously reported results and show that the latter follow as special cases. We end Section III by giving a simple numerical example of wave propagation at a glass-air planar interface, including the effects of total internal reflection. Finally, in Section IV, we summarise and conclude our work.

\section{The Scattering and Transfer Matrices For a Continuous Angular Spectrum}
\subsection{Preliminaries}
We consider the scattering problem depicted in Figure 1. A dielectric scattering medium is situated within the region $-l \leq z \leq l$. We denote by $\mathcal{R}^-$ and $\mathcal{R}^+$ the regions $-L < z < -l$ and $l < z < L$ surrounding the scattering medium, which are assumed to be dielectric with constant permittivity $\epsilon_1$. All sources are contained in the regions $z < -L$ and $z > L$, and may produce both propagating and evanescent incident fields. We assume that the scattering medium is linear and all fields are monochromatic with angular frequency $\omega$. The scattering medium can be described by a spatially inhomogeneous, complex-valued permittivity function $\epsilon(\mathbf{r}, \omega)$, where $\mathbf{r} = (x, y, z)^{\mathrm{T}}$ is the position vector and the superscript $\mathrm{T}$ is used to denote the transpose of a vector or matrix. We also assume that both the scattering and background media are non-magnetic and have magnetic permeabilities equal to the vacuum permeability $\mu_0$. We denote by $\mathbf{E}(\mathbf{r})$ and $\mathbf{H}(\mathbf{r})$ the complex phasor representations of the real electric and magnetic fields with a suppressed time factor of $\exp(-i\omega t)$.

The frequency-domain Maxwell equations for the entire region $-L<z<L$ are given by \cite{mishchenko2006multiple}
\begin{align}
	\nabla \cdot \mathbf{E}(\mathbf{r}) &= 0,\label{ediv}\\
	\nabla \cdot \mathbf{H}(\mathbf{r}) &= 0, \\
	\nabla \times  \mathbf{E}(\mathbf{r}) &= i\omega \mu_0 \mathbf{H}(\mathbf{r}), \label{ecurl}\\
	\nabla \times  \mathbf{H}(\mathbf{r}) &= -i\omega \epsilon(\mathbf{r}, \omega) \mathbf{E}(\mathbf{r}).\label{hcurl}
\end{align} 
From Eqs. (\ref{ecurl}) and (\ref{hcurl}), the vector wave equation for the electric field can be shown to be
\begin{align}\label{eq:wave}
\nabla \times \nabla \times \mathbf{E}(\mathbf{r}) - k^2 \mathbf{E}(\mathbf{r}) = \mathbf{0},
\end{align}
where $k^2 = \omega^2 \mu_0 \epsilon(\mathbf{r},\omega)$.
\begin{figure}\label{test1}
	\includegraphics[scale = 0.4]{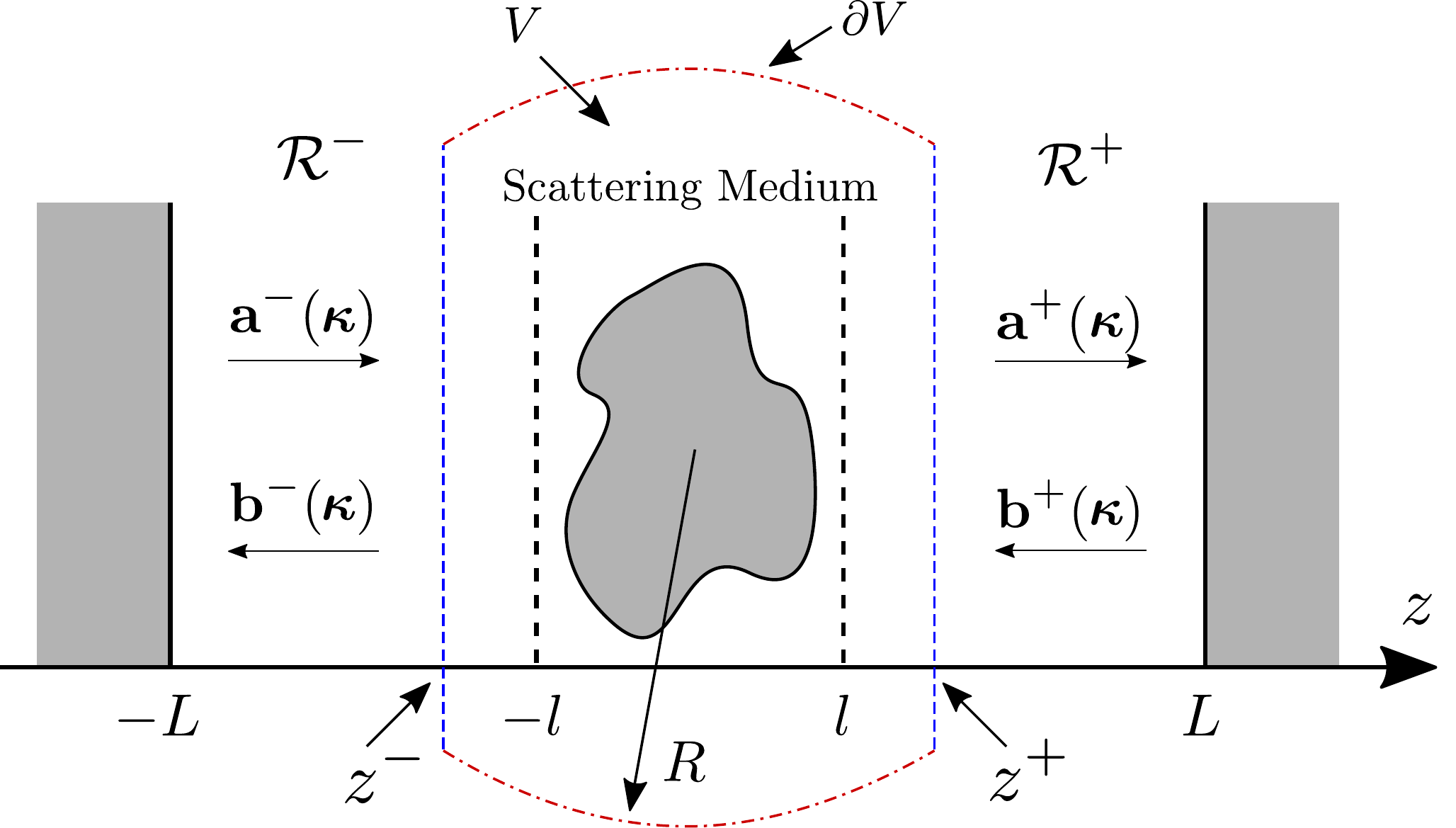}
	\caption{Geometry of the scattering problem. The integrating surface $\partial V$ used in deriving the energy conservation and reciprocity constraints is also depicted. Dashed lines denote the planar sections of $\partial V$ and dot-dashed curves denote the spherical section of radius $R$.}
\end{figure}
In the regions $\mathcal{R}^-$ and $\mathcal{R}^+$, we may write the electric field as a sum of plane wave components using the well-known angular spectrum representation \cite{born1999principles}
\begin{align}\label{eq:angspec}
\begin{split}
\mathbf{E}^{\pm}(\mathbf{r}) &= \int
\mathbf{a}^{\pm}(\bm{\kappa})e^{i(\bm{\kappa}\cdot \bm{\rho} + k_z z)}\mathrm{d}\bm{\kappa} \\
&\quad + \int\mathbf{b}^{\pm}(\bm{\kappa})e^{i(\bm{\kappa}\cdot \bm{\rho} - k_z z)}\mathrm{d}\bm{\kappa},
\end{split}
\end{align}
where the plus or minus sign is chosen according to the region under consideration and $\mathrm{d}\bm{\kappa} = \mathrm{d}k_x\mathrm{d}k_y$. In Eq. (\ref{eq:angspec}) we have introduced the transverse position vector $\bm{\rho} = (x, y)^{\mathrm{T}}$ and transverse wavevector $\bm{\kappa} = (k_x, k_y)^{\mathrm{T}}$. For each plane wave, the $z$ component of the associated wavevector $k_z = k_z(\bm{\kappa})$ is given by
\begin{align}
k_z(\bm{\kappa}) = 
\begin{cases}
\sqrt{k^2 - |\bm{\kappa}|^2} & \text{if $|\bm{\kappa}|^2 \leq k^2$}, \\
i\sqrt{|\bm{\kappa}|^2 - k^2} & \text{if $|\bm{\kappa}|^2 > k^2 $}.
\end{cases}
\end{align}
The vector $\mathbf{a}^{\pm}(\bm{\kappa})$ denotes the amplitude of the right-travelling plane wave with wavevector $\mathbf{k} = (k_x, k_y, k_z)^{\mathrm{T}}$. Similarly, the vector $\mathbf{b}^{\pm}(\bm{\kappa})$ denotes the amplitude of the left-travelling plane wave with wavevector $\widetilde{\mathbf{k}} = (k_x, k_y, -k_z)^{\mathrm{T}}$. In Eq. (\ref{eq:angspec}) and all integrals that follow, the domain of integration is assumed to be from $-\infty$ to $\infty$ for all integration variables unless specified otherwise. The corresponding angular spectrum representation for the magnetic field can be obtained by taking the curl of Eq. (\ref{eq:angspec}) and using Eq. (\ref{ecurl}). It is also necessary that the electric field satisfies the divergence condition in Eq. (\ref{ediv}). By taking the divergence of Eq. (\ref{eq:angspec}) and using Eq. (\ref{ediv}), we find that 
\begin{align}
	\mathbf{a}^{\pm}(\bm{\kappa})\cdot \mathbf{k} = 0,\label{eq:dof1}\\
	\mathbf{b}^{\pm}(\bm{\kappa})\cdot \widetilde{\mathbf{k}} = 0,\label{eq:dof2}
\end{align}
for all $\bm{\kappa}$.  

In $\mathcal{R}^-$ and $\mathcal{R}^+$, $k^2$ is a constant since $k^2 = n^2k_0^2  = n^2(2\pi/\lambda_0)^2$, where $\lambda_0$ is the wavelength in vacuum, $n = \sqrt{\epsilon_1/\epsilon_0}$ is the refractive index and $\epsilon_0$ is the vacuum permittivity. Clearly when $|\bm{\kappa}|^2 \leq k^2$, $k_z$ is real and the plane waves are homogeneous, or propagating. When $|\bm{\kappa}|^2 > k^2 $, $k_z$ is imaginary and the plane waves are inhomogeneous, or evanescent. For convenience, we define the sets $\Gamma_p$ and $\Gamma_e$, where $\Gamma_p = \{\bm{\kappa}:|\bm{\kappa}|^2 \leq k^2\}$ is the set of all transverse wavevectors corresponding to propagating plane waves and $\Gamma_e = \{\bm{\kappa}:|\bm{\kappa}|^2 > k^2\}$ is the set of all transverse wavevectors corresponding to evanescent plane waves.

For linear scattering, we may relate the amplitudes of plane waves travelling towards and away from the scattering medium using the scattering matrix $\mathbf{S}$. Specifically, if we define the column vectors $\mathbf{I}(\bm{\kappa}) = [\mathbf{a}^-(\bm{\kappa}), \mathbf{b}^+(\bm{\kappa})]^{\mathrm{T}}$ and $\mathbf{O}(\bm{\kappa}) = [\mathbf{b}^-(\bm{\kappa}), \mathbf{a}^+(\bm{\kappa})]^{\mathrm{T}}$, then the continuous scattering matrix is defined to be the matrix that satisfies
\begin{align}\label{eq:scatmat}
\mathbf{O}(\bm{\kappa}) = \int \mathbf{S}(\bm{\kappa}, \bm{\kappa}')\mathbf{I}(\bm{\kappa}')\mathrm{d}\bm{\kappa}'.
\end{align}
For a given $\bm{\kappa}$ and $\bm{\kappa}'$, the scattering matrix $\mathbf{S}(\bm{\kappa}, \bm{\kappa}')$ is a $6\times6$ matrix of complex entries, but it is useful to write it as a $2\times 2$ block matrix in the form
\begin{align}\label{eq:scatmatblock}
\mathbf{S}(\bm{\kappa}, \bm{\kappa}') = \begin{pmatrix}
\mathbf{r}(\bm{\kappa}, \bm{\kappa}') & \mathbf{t}'(\bm{\kappa}, \bm{\kappa}')\\
\mathbf{t}(\bm{\kappa}, \bm{\kappa}') & \mathbf{r}'(\bm{\kappa}, \bm{\kappa}')
\end{pmatrix},
\end{align}
where $\mathbf{r}, \mathbf{r}', \mathbf{t}$ and $\mathbf{t}'$ are $3\times 3$ matrix generalizations of transmission and reflection coefficients. 

An alternative description of the scattering problem is possible using the transfer matrix $\mathbf{M}$, which relates the amplitudes of plane waves on the left and right hand side of the medium. Letting $\mathbf{L}(\bm{\kappa}) = [\mathbf{a}^-(\bm{\kappa}), \mathbf{b}^-(\bm{\kappa})]^{\mathrm{T}}$ and $\mathbf{R}(\bm{\kappa}) = [\mathbf{a}^+(\bm{\kappa}), \mathbf{b}^+(\bm{\kappa})]^{\mathrm{T}}$, the continuous transfer matrix is defined to be the matrix that satisfies
\begin{align}\label{eq:transdef}
\mathbf{R}(\bm{\kappa}) = \int \mathbf{M}(\bm{\kappa}, \bm{\kappa}')\mathbf{L}(\bm{\kappa}')\mathrm{d}\bm{\kappa}'.
\end{align}
As with the scattering matrix, it is useful to write the transfer matrix as a $2\times2$ block matrix in the form
\begin{align}\label{tblock}
\mathbf{M}(\bm{\kappa}, \bm{\kappa}') = \begin{pmatrix}
\bm{\alpha}(\bm{\kappa}, \bm{\kappa}') & \bm{\beta}(\bm{\kappa}, \bm{\kappa}')\\
\bm{\gamma}(\bm{\kappa}, \bm{\kappa}') & \bm{\delta}(\bm{\kappa}, \bm{\kappa}')
\end{pmatrix}.
\end{align}
Unlike the scattering matrix, however, the sub-matrices $\bm{\alpha}, \bm{\beta}, \bm{\gamma}$ and $\bm{\delta}$, do not have such an obvious physical interpretation.
\subsection{Conservation of Energy}
We now consider the constraints placed upon the continuous scattering and transfer matrices by energy conservation. In order to enforce conservation of energy, we consider the time-averaged Poynting vector associated with the fields, which is given by $\langle\mathbf{S}(\mathbf{r})\rangle = \frac{1}{2}\mathrm{Re}[\mathbf{E}(\mathbf{r})\times \mathbf{H}^*(\mathbf{r})]$. The average net rate $W$ at which electromagnetic energy flows out of any closed surface $A$ is given by \cite{born1999principles}
\begin{align}\label{eq:coe}
W = \int_{A} \langle\mathbf{S}\rangle\cdot \hat{\mathbf{n}}\,\mathrm{d}A,
\end{align}
where $\hat{\mathbf{n}}$ is the outward unit normal vector to the surface. If energy is not absorbed within the volume enclosed by the surface, then conservation of energy demands that $W = 0$ and the integral in Eq. (\ref{eq:coe}) must vanish. We shall henceforth assume that this is the case.

Consider now the surface  $\partial V$ shown in Figure 1 bounding the volume $V$. The surface consists of several parts: two planar sections at $z = z^-$ and $z = z^+$, and a spherical section of radius $R$ in between the two planar sections. In the limit $R \to \infty$, the two planar sections expand to infinite planes and the energy flow through the spherical section becomes negligible. The total energy flow through the planes can be written as 
\begin{align}\label{eq:energy}
W^+ + W^- = \int_{z = z^+} \!\!\!\!\!\!\! \langle\mathbf{S}\rangle\cdot \hat{\mathbf{z}}\,\mathrm{d}\bm{\rho} - \int_{z = z^-} \!\!\!\!\!\!\!\langle\mathbf{S}\rangle\cdot \hat{\mathbf{z}}\,\mathrm{d}\bm{\rho} = 0,
\end{align} 
where $\mathrm{d}\bm{\rho} = \mathrm{d}x\mathrm{d}y$. The vector $\langle\mathbf{S}\rangle$ can be expressed in terms of the angular spectrums of the electric and magnetic fields defined previously. Doing so and evaluating the  integrals in Eq. (\ref{eq:energy}) yields
\begin{align}\label{eq:coemaster}
\begin{split}
\int_{\Gamma_p} k_z\big[|\bm{a}^-|^2 - |\bm{a}^+|^2 - (|\bm{b}^-|^2 - |\bm{b}^+|^2)\big]\,\mathrm{d}\bm{\kappa}&\\
 + 2i\int_{\Gamma_e} k_z\mathrm{Im}[\bm{a}^-\cdot\bm{b}^{-*} - \bm{a}^+\cdot\bm{b}^{+*}]\,\mathrm{d}\bm{\kappa}& = \,0,
\end{split}
\end{align}
where the dependence of each term on $\bm{\kappa}$ has been temporarily omitted for brevity.

To derive relations pertaining to the scattering matrix, we notice that Eq. (\ref{eq:coemaster}) can be recast in the form
\begin{align}\label{eq:coescatmat}
\begin{split}
&\int_{\Gamma_p}k_z[|\mathbf{I}(\bm{\kappa})|^2 - |\mathbf{O}(\bm{\kappa})|^2]\,\mathrm{d}\bm{\kappa} \\
&+ \int_{\Gamma_e} k_z[\mathbf{I}(\bm{\kappa})\cdot\mathbf{O}^*(\bm{\kappa}) - \mathbf{I}^*(\bm{\kappa})\cdot\mathbf{O}(\bm{\kappa})]\,\mathrm{d}\bm{\kappa} = 0.
\end{split}
\end{align}
Introducing the scattering matrix into Eq. (\ref{eq:coescatmat}) using Eq. (\ref{eq:scatmat}) and simplifying the resultant expressions, we ultimately arrive at
\begin{align}\label{eq:sunit}
\begin{split}
&\int_{\Gamma_p} k_z\mathbf{S}^{\dagger}(\bm{\kappa}, \bm{\kappa}')\mathbf{S}(\bm{\kappa}, \bm{\kappa}'')\,\mathrm{d}\bm{\kappa} \\
&\quad= \begin{cases}
 k_z'\delta(\bm{\kappa}' - \bm{\kappa}'')\mathbb{I}_6 & \text{if $\bm{\kappa}'\in \Gamma_p$,  $\bm{\kappa}''\in \Gamma_p$}, \\
k_z''\mathbf{S}^{\dagger}(\bm{\kappa}'', \bm{\kappa}') & \text{if $\bm{\kappa}'\in \Gamma_p$, $\bm{\kappa}''\in \Gamma_e$}, \\
-k_z'\mathbf{S}(\bm{\kappa}', \bm{\kappa}'') & \text{if $\bm{\kappa}'\in \Gamma_e$, $\bm{\kappa}''\in \Gamma_p$}, \\
k_z''\mathbf{S}^{\dagger}(\bm{\kappa}'', \bm{\kappa}') - k_z'\mathbf{S}(\bm{\kappa}', \bm{\kappa}'') & \text{if $\bm{\kappa}'\in \Gamma_e$, $\bm{\kappa}''\in \Gamma_e$}, 
\end{cases}
\end{split}
\end{align}
where $\delta$ is the Dirac delta function, the superscript $\dagger$ denotes the conjugate transpose and we use $\mathbb{I}_n$ to denote the $n\times n$ identity matrix. These so-called extended unitarity relations are hence an expression of conservation of energy and generalize the better known unitarity condition on the scattering matrix so as to include evanescent wave components \cite{PhysRevA.62.012712}. 

The transfer matrix can also be shown to obey a similar set of equations to Eq. (\ref{eq:sunit}).  Returning to Eq. (\ref{eq:coemaster}), we note that the integrands can be written in the alternate form
\begin{align}\label{eq:transunit}
\begin{split}
&\int_{\Gamma_p}k_z[\mathbf{L}^{\dagger}(\bm{\kappa})\bm{\Sigma}^z_{3}\mathbf{L}(\bm{\kappa}) - \mathbf{R}^{\dagger}(\bm{\kappa})\bm{\Sigma}^z_{3}\mathbf{R}(\bm{\kappa})]\,\mathrm{d}\bm{\kappa} \\
&- \int_{\Gamma_e}ik_z[\mathbf{L}^{\dagger}(\bm{\kappa})\bm{\Sigma}^y_{3}\mathbf{L}(\bm{\kappa}) - \mathbf{R}^{\dagger}(\bm{\kappa})\bm{\Sigma}^y_{3}\mathbf{R}(\bm{\kappa})]\,\mathrm{d}\bm{\kappa} = 0,
\end{split}
\end{align}
where we have introduced the generalized Pauli matrices
\begin{align}
\bm{\Sigma}^z_{n} = \begin{pmatrix}
\mathbb{I}_n &\mathbb{O}_n \\
\mathbb{O}_n &-\mathbb{I}_n
\end{pmatrix},\quad
\mathbf{\Sigma}^y_{n} = i\begin{pmatrix}
\mathbb{O}_n  &-\mathbb{I}_n \\
\mathbb{I}_n &\mathbb{O}_n
\end{pmatrix},
\end{align}
where $\mathbb{O}_n$ denotes the $n \times n$ zero matrix. Similarly to before, we introduce the transfer matrix into Eq. (\ref{eq:transunit}) using Eq. (\ref{eq:transdef}), which yields 
\begin{align}
\begin{split}\label{eq:munit}
&\int_{\Gamma_p}k_z\mathbf{M}^{\dagger}(\bm{\kappa}, \bm{\kappa}'')\bm{\Sigma}^z_{3}\mathbf{M}(\bm{\kappa}, \bm{\kappa}')\,\mathrm{d}\bm{\kappa} \\
 &-  \int_{\Gamma_e}ik_z\mathbf{M}^{\dagger}(\bm{\kappa}, \bm{\kappa}'')\bm{\Sigma}^y_{3}\mathbf{M}(\bm{\kappa}, \bm{\kappa}')\,\mathrm{d}\bm{\kappa} \\
 &\quad= \begin{cases}
 k_z'\delta(\bm{\kappa}' - \bm{\kappa}'')\bm{\Sigma}^z_{3} & \text{if $\bm{\kappa}'\in \Gamma_p$,  $\bm{\kappa}''\in \Gamma_p$}, \\
 -ik_z' \delta(\bm{\kappa}' - \bm{\kappa}'')\bm{\Sigma}^y_{3} & \text{if $\bm{\kappa}'\in \Gamma_e$,  $\bm{\kappa}''\in \Gamma_e$}, \\
 \mathbb{O}_6 & \text{otherwise.}
 \end{cases}
\end{split}
\end{align}
Eq. (\ref{eq:munit}) therefore represents the constraint imposed upon the transfer matrix by energy conservation.
\subsection{Reciprocity}
In this section, we consider the constraints imposed upon the scattering and transfer matrices due to the reciprocity principle. Roughly speaking, reciprocity describes a relation between scattering matrices that are related by an exchange of input and output modes. For a more detailed review, see  Ref. \cite{10.1088/0034-4885/67/5/r03}.

Consider again the volume $V$ shown in Figure 1 and its bounding surface $\partial V$. Let $\mathbf{E}_1$ and $\mathbf{E}_2$ be two arbitrary fields that satisfy Maxwell's equations. Applying the vector analogue of Green's second identity to these two fields gives \cite{PhysRev.56.99}
\begin{align}\label{eq:greens}
\begin{split}
&\int_{V}(\mathbf{E}_1\cdot\nabla\times\nabla\times \mathbf{E}_2 - \mathbf{E}_2\cdot\nabla\times\nabla\times \mathbf{E}_1)\,\mathrm{d}V \\
&\quad= \int_{\partial V}(\mathbf{E}_2\times\nabla\times\mathbf{E}_1 - \mathbf{E}_1\times\nabla\times\mathbf{E}_2)\cdot\hat{\mathbf{n}}\,\mathrm{d}A.
\end{split}
\end{align}
Since both fields satisfy Eq. (\ref{eq:wave}) in $V$, it can be shown that the integral on the left hand side of Eq. (\ref{eq:greens}) vanishes. Furthermore, by writing the electric fields in the far field as sums of incoming and outgoing spherical waves (see e.g. Ref. \cite{mishchenko2006multiple}), it can be shown that the integral on the right hand side of Eq. (\ref{eq:greens}) over the spherical section of $\partial V$ vanishes in the limit $R \to \infty$. Therefore, we conclude that the integral on the right hand side of Eq. (\ref{eq:greens}) over the infinite planes $z = z^-$ and $z = z^+$ is equal to zero. Expressing $\mathbf{E}_1$ and $\mathbf{E}_2$ using angular spectrum representations, we arrive at the equation $I_{z^-} = I_{z^+}$, where
\begin{align}\label{eq:recmaster}
\begin{split}
I_{z^{\pm}} = \int k_z[&\mathbf{a}^{\pm}_1(\bm{\kappa})\cdot\mathbf{b}^{\pm}_2(-\bm{\kappa}) + \mathbf{a}^{\pm}_1(-\bm{\kappa})\cdot\mathbf{b}^{\pm}_2(\bm{\kappa}) \\
- &\mathbf{a}^{\pm}_2(\bm{\kappa})\cdot\mathbf{b}^{\pm}_1(-\bm{\kappa}) - \mathbf{a}^{\pm}_2(-\bm{\kappa})\cdot\mathbf{b}^{\pm}_1(\bm{\kappa})]\mathrm{d}\bm{\kappa}.
\end{split}
\end{align}
Written in terms of the vectors $\mathbf{I}$ and $\mathbf{O}$, the equation $I_{z^-} = I_{z^+}$ becomes
\begin{align}\label{eq:rs}
\int k_z[\mathbf{I}_1(\bm{\kappa})\cdot \mathbf{O}_2(-\bm{\kappa}) - \mathbf{I}_2(\bm{\kappa})\cdot\mathbf{O}_1(-\bm{\kappa})]\mathrm{d}\bm{\kappa} = 0.
\end{align}
By now introducing the scattering matrix into Eq. (\ref{eq:rs}) using Eq. (\ref{eq:scatmat}), we are able to derive the reciprocity relation for the scattering matrix
\begin{align}\label{eq:srec}
k_z(\bm{\kappa}')\mathbf{S}(\bm{\kappa}', \bm{\kappa}'') = k_z(\bm{\kappa}'')\mathbf{S}^{\mathrm{T}}(-\bm{\kappa}'', -\bm{\kappa}'),
\end{align}
which is valid for all pairs of transverse wavevectors $\bm{\kappa}'$ and $\bm{\kappa}''$. Reciprocity relations for the constituent transmission and reflection matrices can be obtained by considering the block form of the scattering matrix as in Eq. (\ref{eq:scatmatblock}). Comparing blocks on either side of Eq. (\ref{eq:srec}), we obtain 
\begin{align}
k_z(\bm{\kappa}')\mathbf{r}(\bm{\kappa}', \bm{\kappa}'') &= k_z(\bm{\kappa}'')\mathbf{r}^{\mathrm{T}}(-\bm{\kappa}'', -\bm{\kappa}'),\\
k_z(\bm{\kappa}')\mathbf{r}'(\bm{\kappa}', \bm{\kappa}'') &= k_z(\bm{\kappa}'')\mathbf{r}'^{\mathrm{T}}(-\bm{\kappa}'', -\bm{\kappa}'),\\
k_z(\bm{\kappa}')\mathbf{t}(\bm{\kappa}', \bm{\kappa}'') &= k_z(\bm{\kappa}'')\mathbf{t}'^{\mathrm{T}}(-\bm{\kappa}'', -\bm{\kappa}'),
\end{align}
which are consistent with those previously reported \cite{Carminati:98}.

Alternatively, writing Eq. (\ref{eq:recmaster}) in terms of the vectors $\mathbf{L}$ and $\mathbf{R}$, we find
\begin{align}
\int k_z[\mathbf{L}^{\mathrm{T}}_1(\bm{\kappa})\bm{\Sigma}^y_{3}\mathbf{L}_2(-\bm{\kappa}) - \mathbf{R}^{\mathrm{T}}_1(\bm{\kappa})\bm{\Sigma}^y_{3}\mathbf{R}_2(-\bm{\kappa})]\mathrm{d}\bm{\kappa} = 0,
\end{align}
which, after introducing the transfer matrix using Eq. (\ref{eq:transdef}), gives
\begin{align}\label{eq:transrec}
\int k_z\mathbf{M}^{\mathrm{T}}(\bm{\kappa}, \bm{\kappa}')\bm{\Sigma}^y_{3}\mathbf{M}(-\bm{\kappa}, \bm{\kappa}'')\mathrm{d}\bm{\kappa} = k_z'\delta(\bm{\kappa}'+\bm{\kappa}'')\bm{\Sigma}^y_{3}.
\end{align}
Eq. (\ref{eq:transrec}) is thus the reciprocity relation for the transfer matrix.
\subsection{Time Reversal Symmetry}
Let us now consider time reversal symmetry, which describes an invariance under the transformation $t \to -t$. Let $\mathbf{E}_1$ be an arbitrary, real electric field and let $\mathbf{E}_2$ be its time reversed counterpart, i.e. $\mathbf{E}_2(\mathbf{r}, t) = \mathbf{E}_1(\mathbf{r}, -t)$. It can be shown that the Fourier transforms of the fields satisfy $\mathbf{E}_2(\mathbf{r}, \omega) = \mathbf{E}^*_1(\mathbf{r}, \omega)$ \cite{Leuchs_2012}. Making use of this property and comparing the angular spectrums of the two fields, we find
\begin{align}
&\mathbf{a}^{\pm}_2(\bm{\kappa}) = \mathbf{b}^{\pm *}_1(-\bm{\kappa}),\label{st1} \\
&\mathbf{b}^{\pm}_2(\bm{\kappa})= \mathbf{a}^{\pm *}_1(-\bm{\kappa}), \
\end{align}
for $\bm{\kappa} \in \Gamma_p$ and
\begin{align}
&\mathbf{a}^{\pm}_2(\bm{\kappa})= \mathbf{a}^{\pm *}_1(-\bm{\kappa}), \\
&\mathbf{b}^{\pm}_2(\bm{\kappa}) = \mathbf{b}^{\pm *}_1(-\bm{\kappa}), \label{st2}\
\end{align}
for $\bm{\kappa} \in \Gamma_e$. In terms of the scattering matrix, a system is time reversal invariant if both $\mathbf{E}_1$ and $\mathbf{E}_2$ satisfy Eq. (\ref{eq:scatmat}) for the same scattering matrix $\mathbf{S}$. Combining this property with Eqs. (\ref{st1})-(\ref{st2}), we ultimately find 
\begin{align}\label{eq:strs}
\begin{split}
&\int_{\Gamma_p}\mathbf{S}(\bm{\kappa}', \bm{\kappa})\mathbf{S}^{*}(-\bm{\kappa}, \bm{\kappa}'')\,\mathrm{d}\bm{\kappa} \\
&\quad= \begin{cases}
\delta(\bm{\kappa}' + \bm{\kappa}'')\mathbb{I}_6 & \text{if $\bm{\kappa}'\in \Gamma_p$,  $\bm{\kappa}''\in \Gamma_p$}, \\
-\mathbf{S}(\bm{\kappa}', -\bm{\kappa}'') & \text{if $\bm{\kappa}'\in \Gamma_p$, $\bm{\kappa}''\in \Gamma_e$}, \\
\mathbf{S}^*(-\bm{\kappa}', \bm{\kappa}'') & \text{if $\bm{\kappa}'\in \Gamma_e$, $\bm{\kappa}''\in \Gamma_p$}, \\
- \mathbf{S}(\bm{\kappa}', -\bm{\kappa}'')+\mathbf{S}^{*}(-\bm{\kappa}', \bm{\kappa}'') & \text{if $\bm{\kappa}'\in \Gamma_e$, $\bm{\kappa}''\in \Gamma_e$}, 
\end{cases}
\end{split}
\end{align}
which is therefore an expression of time reversal symmetry for the scattering matrix.

If we instead require that both fields satisfy Eq. (\ref{eq:transdef}) for the same transfer matrix, we find
\begin{align}
\begin{split}\label{trstrans}
&\mathbf{M}(\bm{\kappa}', \bm{\kappa}'')\\ &\quad= \begin{cases}
\bm{\Sigma}^x_{3} \mathbf{M}^*(-\bm{\kappa}', -\bm{\kappa}'')\bm{\Sigma}^x_{3} & \text{if $\bm{\kappa}'\in \Gamma_p$,  $\bm{\kappa}''\in \Gamma_p$}, \\
\bm{\Sigma}^x_{3} \mathbf{M}^*(-\bm{\kappa}', -\bm{\kappa}'') & \text{if $\bm{\kappa}'\in \Gamma_p$, $\bm{\kappa}''\in \Gamma_e$}, \\
\mathbf{M}^*(-\bm{\kappa}', -\bm{\kappa}'')\bm{\Sigma}^x_{3} & \text{if $\bm{\kappa}'\in \Gamma_e$, $\bm{\kappa}''\in \Gamma_p$}, \\
\mathbf{M}^*(-\bm{\kappa}', -\bm{\kappa}'') & \text{if $\bm{\kappa}'\in \Gamma_e$, $\bm{\kappa}''\in \Gamma_e$}, 
\end{cases}
\end{split}
\end{align}
where
\begin{align}
\mathbf{\Sigma}^x_{n} = \begin{pmatrix}
\mathbb{O}_n & \mathbb{I}_n \\
\mathbb{I}_n & \mathbb{O}_n\ 
\end{pmatrix}.
\end{align}
Eq. (\ref{trstrans}) is therefore the time reversal symmetry constraint for the transfer matrix.

To conclude this section, we note briefly that for the scattering matrix the time reversal symmetry equations also follow from the conservation of energy and reciprocity equations, i.e. Eq. (\ref{eq:sunit}) and Eq. (\ref{eq:srec}) imply Eq. (\ref{eq:strs}). Similarly, for the transfer matrix, the reciprocity relation can be derived from conservation of energy and time reversal symmetry, i.e. Eq. (\ref{eq:munit}) and Eq. (\ref{trstrans}) imply Eq. (\ref{eq:transrec}). We therefore conclude that, in accordance with Ref. \cite{PhysRevA.62.012712}, for a system that conserves energy, reciprocity and time reversal symmetry are equivalent. 

\section{The Scattering and Transfer Matrices For A Discrete Angular Spectrum}
The scattering and transfer matrices we have considered so far are, in principle, defined for all pairs of conceivable transverse wavevectors $\bm{\kappa}'$ and $\bm{\kappa}''$, which form a continuous spectrum and are infinite in number. In both experiments and numerical simulations, however, fields cannot be resolved with infinite precision and must be described using some finite set of modes \cite{Miller:19}. Beyond merely being a practical limitation, the number of independent modes a system can support in reality must be finite due to the wave nature of light and is often constrained by the geometry of the scattering system. This is particularly relevant for waveguides, such as optical fibers, where the number of modes is finite and is determined by the fiber's radius and refractive indices \cite{Sethi18}. Even for waves in free space, however, diffraction places a lower limit on the resolution to which a field can be discretely sampled \cite{Judkewitz2015}. Sampling beyond this limit would result in a set of modes that would not be independent and would therefore not yield additional information.

In this section, we shall consider a pixel-wise discretization of the continuous spectrum of transverse wavevectors. This will allow us to define scattering and transfer matrices that describe coupling between plane waves in a discrete angular spectrum. We shall then derive the constraints that must be satisfied by these matrices.

\subsection{Definitions}
Let us consider a finite set of plane waves indexed by their transverse wavevectors. We first define two sets, $K_p$ and $K_e$, of transverse wavevectors for propagating and evanescent waves respectively. We form $K_p$ by choosing $N_p$ transverse wavevectors $\bm{\kappa}^p_i$ corresponding to propagating waves together with their additive inverses $-\bm{\kappa}^p_i$ and possibly the zero vector. As we shall demonstrate, it is necessary to include modes in inverse pairs to fully explore the effects of reciprocity and time reversal symmetry. Similarly, we construct $K_e$ by taking $N_e$ transverse wavevectors $\bm{\kappa}^e_i$ corresponding to evanescent waves and their additive inverses $-\bm{\kappa}^e_i$. Thus, we have
\begin{align}
		&K_p = \{ -\bm{\kappa}^{p}_{N_{p}}, \ldots, -\bm{\kappa}^{p}_{2}, -\bm{\kappa}^{p}_{1}, \bm{0}, 	\bm{\kappa}^{p}_{1},  \bm{\kappa}^{p}_{2}, \ldots, \bm{\kappa}^{p}_{N_{p}} \}, \label{modes1}\\
		&K_e = \{ -\bm{\kappa}^{e}_{N_{e}}, \ldots, -\bm{\kappa}^{e}_{2}, -\bm{\kappa}^{e}_{1}, \bm{\kappa}^{e}_{1},  \bm{\kappa}^{e}_{2}, \ldots, \bm{\kappa}^{e}_{N_{e}} \}, \label{modes2}\	
\end{align}
which, with the inclusion of the zero vector in $K_p$, contain $2N_p+1$ and $2N_e$ elements respectively. The set of all wavevectors $K$ is then the union of these sets, i.e. $K = K_p \cup K_e$. A sample of several modes is shown in Figure 2. Note that while in Figure 2 we have, for simplicity, distributed the modes at points on a rectangular lattice in $k$-space, the choice of other geometries, such as a hexagonal lattice, may have practical advantages \cite{Pai:20}. Note also that although $k$-space is unbounded, there is a practical upper limit to the size of $|\bm{\kappa}^e_i|$. This is because when $|\bm{\kappa}^e_i|$ is sufficiently large, its corresponding wave amplitude, even at positions very close to the scattering medium, will have decayed to the point of being practically unmeasurable.
\begin{figure}
	\includegraphics[scale = 0.45]{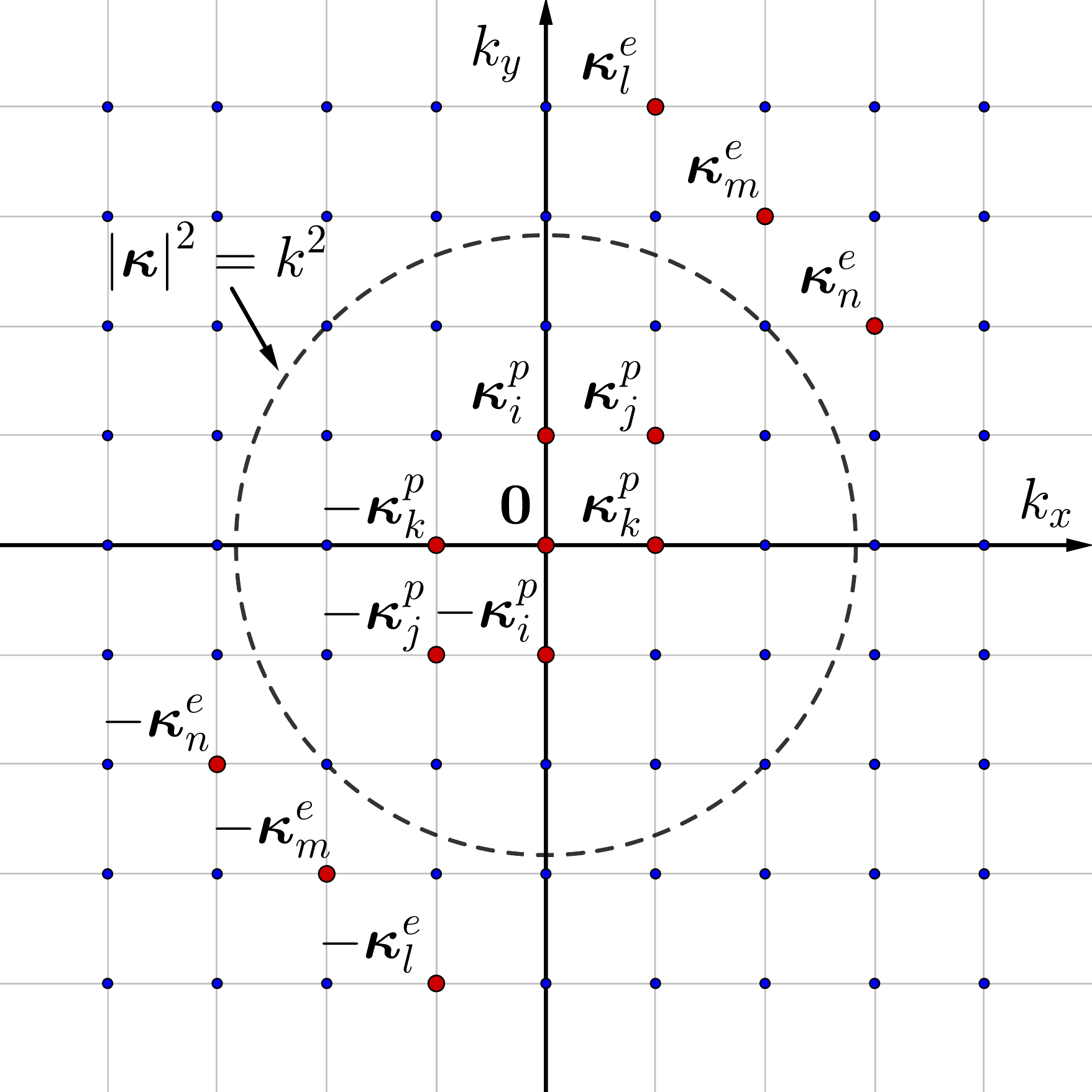}
	\label{fig:modes}
	\caption{Transverse wavevector modes distributed on a rectangular lattice in $k$-space. The circle $|\bm\kappa|^2 = k^2$ defines the boundary between propagating and evanescent modes. A selection of modes and their additive inverses have been highlighted in red.}
\end{figure}

As a result of Eqs. (\ref{eq:dof1}) and (\ref{eq:dof2}), the amplitude associated with each plane wave has only two degrees of freedom. Therefore, in order to remove extraneous information and simplify the ensuing mathematics, it is useful to introduce the vectors
\begin{align}
		&\mathbf{e}_k(\bm{\kappa}, k_z) = \frac{\mathbf{k}}{k}= \frac{1}{k}\begin{pmatrix}
			k_x \\ k_y \\ k_z \
		\end{pmatrix},  \\	
		&\mathbf{e}_\phi(\bm{\kappa}, k_z) = \frac{(\hat{\mathbf{z}} \times  \mathbf{e}_k)}{|\hat{\mathbf{z}} \times  \mathbf{e}_k|} = \frac{1}{|\bm{\kappa}|}\begin{pmatrix}
			-k_y \\ k_x  \\ 0 \
		\end{pmatrix}, \\
		&\mathbf{e}_\theta(\bm{\kappa}, k_z) =  \frac{(\mathbf{e}_\phi \times \mathbf{e}_k)}{|\mathbf{e}_\phi \times \mathbf{e}_k|}= \frac{1}{k |\bm{\kappa}|}  
		\begin{pmatrix}
			k_xk_z \\ k_yk_z\\  -k_x^2 - k_y^2 \
		\end{pmatrix}.\
\end{align}
Note that when $\bm{\kappa} \in \Gamma_p$, $k_z$ is real and the vectors $\mathbf{e}_k, \mathbf{e}_\phi$ and $\mathbf{e}_\theta$ are the standard unit basis vectors in spherical polar coordinates. When $\bm{\kappa} \in \Gamma_{e}$, $\mathbf{e}_k$ and $\mathbf{e}_\theta$ become complex vectors and the latter may be interpreted using a complex polar angle. The vectors $\mathbf{e}_\phi$ and $\mathbf{e}_\theta$ are also classically referred to as $s$ and $p$ modes in polarisation theory \cite{born1999principles}, but we have chosen to reserve the letter $p$ in this work for `propagating'. By definition, it follows that
\begin{align}
 &\mathbf{e}_\phi(\bm{\kappa}, k_z) \cdot \mathbf{k} = \mathbf{e}_\theta(\bm{\kappa}, k_z) \cdot \mathbf{k} = 0, \\ 
 &\mathbf{e}_\phi(\bm{\kappa}, -k_z) \cdot \widetilde{\mathbf{k}} = \mathbf{e}_\theta(\bm{\kappa}, -k_z) \cdot \widetilde{\mathbf{k}} = 0, \
\end{align}
which means we may express $\mathbf{a}^{\pm}(\bm{\kappa})$ and $\mathbf{b}^{\pm}(\bm{\kappa})$ in Eq. (\ref{eq:angspec}) in terms of their $\theta$ and $\phi$ components. Explicitly, we have
\begin{align}
		&\mathbf{a}^\pm(\bm{\kappa}) = a^{\pm}_{\theta}(\bm{\kappa})\mathbf{e}_\theta(\bm{\kappa}, k_z) + a^{\pm}_{\phi}(\bm{\kappa})\mathbf{e}_\phi(\bm{\kappa}, k_z),\\
		&\mathbf{b}^\pm(\bm{\kappa}) = b^{\pm}_{\theta}(\bm{\kappa})\mathbf{e}_\theta(\bm{\kappa}, -k_z) + b^{\pm}_{\phi}(\bm{\kappa})\mathbf{e}_\phi(\bm{\kappa}, -k_z).\
\end{align}

It is now possible to reduce the matrices $\mathbf{r}, \mathbf{r}', \mathbf{t}, \mathbf{t}', \bm{\alpha},\bm{\beta}, \bm{\gamma}$ and $\bm{\delta}$ in Eqs. (\ref{eq:scatmatblock}) and (\ref{tblock}) to $2\times 2$ matrices that couple the $\theta$ and $\phi$ components of  $\mathbf{a}^{\pm}(\bm{\kappa})$ and  $\mathbf{b}^{\pm}(\bm{\kappa})$. As an example, consider $\mathbf{r}(\bm{\kappa}_i, \bm{\kappa}_j)$, where $\bm{\kappa}_i$ and $\bm{\kappa}_j$ are any two vectors taken from $K$. This matrix describes the reflection at the left side of the medium of an incident wave with wavevector $(\bm{\kappa}_j, k_{zj})^{\mathrm{T}}$ and amplitude $\mathbf{a}^-(\bm{\kappa}_j)$ to a final wave with wavevector $(\bm{\kappa}_i, -k_{zi})^{\mathrm{T}}$ and amplitude $\mathbf{b}^-(\bm{\kappa}_i)$, where $k_{zj} = k_z(\bm\kappa_j)$. We define the $2\times 2$ reduced reflection matrix for the pair of modes $\bm\kappa_i$ and $\bm\kappa_j$ as the matrix $\bm{r}_{(i,j)}$, where
\begin{align}\label{rred}
	\bm{r}_{(i,j)} = \begin{pmatrix}
		r_{(i,j)\theta \theta}&r_{(i,j)\theta \phi}\\
		r_{(i,j)\phi \theta}&r_{(i,j)\phi\phi}\
	\end{pmatrix}
\end{align}
and
\begin{align}\label{app}
		r_{(i,j)mn} = \mathbf{e}^{\mathrm{T}}_{m}(\bm{\kappa}_i, -k_{zi}) \mathbf{r}(\bm{\kappa}_i, \bm{\kappa}_j) \mathbf{e}_{n}(\bm{\kappa}_j, k_{zj}),
\end{align}
where $m$ and $n$ may be chosen to be either $\theta$ or $\phi$. Note that 
\begin{align} 
\mathbf{e}^{\mathrm{T}}_{k}(\bm{\kappa}_i, -k_{zi}) \mathbf{r}(\bm{\kappa}_i, \bm{\kappa}_j) = \mathbf{0}^{\mathrm{T}}
\end{align}
for all $\bm{\kappa}_i$ and $\bm{\kappa}_j$, but, since $\mathbf{a}^-(\bm{\kappa}_j)$ has no $k$ component, $\mathbf{r}(\bm{\kappa}_i, \bm{\kappa}_j)\mathbf{e}_{k}(\bm{\kappa}_j, k_{zj})$ is undefined. We therefore assign
\begin{align} 
	 \mathbf{r}(\bm{\kappa}_i, \bm{\kappa}_j)\mathbf{e}_{k}(\bm{\kappa}_j, k_{zj}) = \mathbf{0},
\end{align}
which justifies the construction of the reduced reflection matrix as only the four components in Eq. (\ref{rred}) are unconstrained. The other sub-matrices of $\mathbf{S}$ and $\mathbf{M}$ can be treated similarly and a brief summary is given in Appendix A.

We can now construct the discrete scattering matrix for our finite set of modes by considering a discretized version of Eq. (\ref{eq:scatmat}). To achieve this, we partition $k$-space into a series of regions, each of which is centred on a mode in the set $K$. We denote the area of these regions by $\Delta \kappa$, which, for convenience, we assume is the same for each region. If, for example, a rectangular partitioning is used, then $\Delta\kappa = \Delta k_x \Delta k_y$, where $\Delta k_x$ and $\Delta k_y$ are the distances in $k$-space between adjacent modes. We assume that the choice of modes and $k$-space partitioning are such that the scattering and transfer matrices are approximately constant over each region. 

Let $\bm{\kappa}_i \in K$ be any transverse wavevector. We may replace the integral in Eq. (\ref{eq:scatmat}) with a sum and write
\begin{align}\label{eq:smd}
	\begin{pmatrix}
		b^{-}_{\theta}(\bm{\kappa}_i)\\
		b^{-}_{\phi}(\bm{\kappa}_i)\\
		a^{+}_{\theta}(\bm{\kappa}_i)\\
		a^{+}_{\phi}(\bm{\kappa}_i)\\
	\end{pmatrix} = \sum_{\bm{\kappa}_j \in K}
\begin{pmatrix}
	\mathbf{r}_{(i,j)} & \mathbf{t}'_{(i,j)}\\
	\mathbf{t}_{(i,j)}& \mathbf{r}'_{(i,j)}
\end{pmatrix}
 	\begin{pmatrix}
	a^{-}_{\theta}(\bm{\kappa}_j)\\
	a^{-}_{\phi}(\bm{\kappa}_j)\\
	b^{+}_{\theta}(\bm{\kappa}_j)\\
	b^{+}_{\phi}(\bm{\kappa}_j)\\
\end{pmatrix}\Delta \kappa,
\end{align}
where $\bm{\kappa}_j$ in the sum ranges over all modes in $K$. By now letting $\bm{\kappa}_i$ vary over the set $K$, we obtain a system of equations, each of which have the same form as Eq. (\ref{eq:smd}), which can be combined into a single matrix equation. To facilitate this, we first introduce the notation
\begin{align}
	\begin{split}\label{unotation}
		\mathbf{u}^{\pm}_{q} = (u^{\pm}_{\theta}(-\bm{\kappa}^{q}_{N_{q}}),u^{\pm}_{\phi}(-\bm{\kappa}^{q}_{N_{q}}),\hdots,u^{\pm}_{\theta}(\bm{\kappa}^{q}_{N_{q}}),u^{\pm}_{\phi}(\bm{\kappa}^{q}_{N_{q}})),
	\end{split}
\end{align}
where $\mathbf{u}$ denotes either $\mathbf{a}$ or $\mathbf{b}$ and $q$ is either $p$ or $e$. Regardless of the choice of $q$, we order the transverse wavevectors within $\mathbf{u}$ from left to right in the same way as they are presented in Eqs. (\ref{modes1}) and (\ref{modes2}). Using the notation in Eq. (\ref{unotation}), we define the four vectors
\begin{align}
		\mathbf{c}_i = (\mathbf{a}^{-}_{p}, \mathbf{b}^{+}_{p}, \mathbf{a}^{-}_{e}, \mathbf{b}^{+}_{e})^\mathrm{T},\\
		\mathbf{c}_o = (\mathbf{b}^{-}_{p}, \mathbf{a}^{+}_{p}, \mathbf{b}^{-}_{e}, \mathbf{a}^{+}_{e})^\mathrm{T},\\
		\mathbf{c}_l = (\mathbf{a}^{-}_{p}, \mathbf{b}^{-}_{p}, \mathbf{a}^{-}_{e}, \mathbf{b}^{-}_{e})^\mathrm{T},\\
		\mathbf{c}_r = (\mathbf{a}^{+}_{p}, \mathbf{b}^{+}_{p}, \mathbf{a}^{+}_{e}, \mathbf{b}^{+}_{e})^\mathrm{T},
\end{align}
with which we define the discrete scattering matrix $\mathbf{S}$ as the matrix satisfying 
\begin{align}\label{s1}
\mathbf{c}_{o} = \mathbf{S}\mathbf{c}_{i}.
\end{align}
Given our ordering of the modes, we may partition the matrix $\mathbf{S}$ in the block form
\begin{align}\label{sblock}
	\mathbf{S} = 
	\begin{pmatrix}
		\mathbf{S}_{pp} &\mathbf{S}_{pe} \\
		\mathbf{S}_{ep} &\mathbf{S}_{ee} \
	\end{pmatrix} = 
\begin{pmatrix}
	\mathbf{r}_{pp}&\mathbf{t}'_{pp}&\mathbf{r}_{pe}&\mathbf{t}'_{pe} \\
	\mathbf{t}_{pp}&\mathbf{r}'_{pp}&\mathbf{t}_{pe}&\mathbf{r}'_{pe} \\
	\mathbf{r}_{ep}&\mathbf{t}'_{ep}&\mathbf{r}_{ee}&\mathbf{t}'_{ee} \\
	\mathbf{t}_{ep}&\mathbf{r}'_{ep}&\mathbf{t}_{ee}&\mathbf{r}'_{ee} \
\end{pmatrix},
\end{align}
where
\begin{align}
	\mathbf{S}_{q q'} = 
	\begin{pmatrix}
	\mathbf{r}_{qq'} &\mathbf{t}'_{qq'} \\
	\mathbf{t}_{qq'} &\mathbf{r}'_{qq'} \
	\end{pmatrix}
\end{align}
and $q$ and $q'$ are either $p$ or $e$. For each sub-matrix, the right subscript denotes the type of incident mode (i.e. propagating or evanescent) and the left subscript denotes the type of outgoing mode. For example, $\mathbf{r}_{pe}$ describes the reflection at the left hand side of the system of incoming evanescent modes to outgoing propagating modes. It is formed by concatenating $2\times 2$ reduced reflection matrices of the form in Eq. (\ref{rred}). All other sub-matrices of $\mathbf{S}$ can be understood in an analogous manner.

Similarly, we define the discrete transfer matrix $\mathbf{M}$ to be the matrix that satisfies 
\begin{align}\label{m111}
\mathbf{c}_{r} = \mathbf{M}\mathbf{c}_{l}.
\end{align}
This matrix can also be partitioned in an analogous way to the discrete scattering matrix. Explicitly, 
\begin{align}\label{mblock}
	\mathbf{M} = 
	\begin{pmatrix}
		\mathbf{M}_{pp} &\mathbf{M}_{pe} \\
		\mathbf{M}_{ep} &\mathbf{M}_{ee} \
	\end{pmatrix} = 
	\begin{pmatrix}
		\bm{\alpha}_{pp}&\bm{\beta}_{pp}&\bm{\alpha}_{pe}&\bm{\beta}_{pe} \\
		\bm{\gamma}_{pp}&\bm{\delta}_{pp}&\bm{\gamma}_{pe}&\bm{\delta}_{pe} \\
		\bm{\alpha}_{ep}&\bm{\beta}_{ep}&\bm{\alpha}_{ee}&\bm{\beta}_{ee} \\
		\bm{\gamma}_{ep}&\bm{\delta}_{ep}&\bm{\gamma}_{ee}&\bm{\delta}_{ee} \
	\end{pmatrix}.
\end{align}
In order to simplify the equations in the following section, it is useful to normalize the scattering and transfer matrices. We use a bar to indicate the normalized version of a matrix and define the normalized continuous scattering and transfer matrices by
\begin{align}
		&\bar{\mathbf{S}}(\bm{\kappa}, \bm{\kappa}') = \frac{\sqrt{|k_z(\bm{\kappa})|}}{\sqrt{|k_z(\bm{\kappa}')|}}\mathbf{S}(\bm{\kappa}, \bm{\kappa}')\Delta \kappa,\label{norm1}\\
		&\bar{\mathbf{M}}(\bm{\kappa}, \bm{\kappa}') = \frac{\sqrt{|k_z(\bm{\kappa})|}}{\sqrt{|k_z(\bm{\kappa}')|}}\mathbf{M}(\bm{\kappa}, \bm{\kappa}')\Delta \kappa.\label{norm2}\
\end{align}
The corresponding normalized discrete scattering and transfer matrices are given by
\begin{align}
		\bar{\mathbf{S}} &= \bm\eta^{\frac{1}{2}} \mathbf{S} \bm\eta^{-\frac{1}{2}}\Delta \kappa,\\
		\bar{\mathbf{M}} &= \bm\eta^{\frac{1}{2}} \mathbf{M} \bm\eta^{-\frac{1}{2}}\Delta \kappa,\
\end{align}
where 
\begin{align}
&\bm{\eta}^{\pm\frac{1}{2}} = \textrm{diag}(\bm{\eta}^{\pm\frac{1}{2}}_p,\bm{\eta}^{\pm\frac{1}{2}}_p,\bm{\eta}^{\pm\frac{1}{2}}_e,\bm{\eta}^{\pm\frac{1}{2}}_e), \\
&\bm{\eta}^{\pm\frac{1}{2}}_p = \textrm{diag}(|k_z(-\bm{\kappa}^{p}_{N_p})|^{\pm\frac{1}{2}}\mathbb{I}_2, \hdots, |k_z(\bm{\kappa}^{p}_{N_p})|^{\pm\frac{1}{2}}\mathbb{I}_2),\\
&\bm{\eta}^{\pm\frac{1}{2}}_e = \textrm{diag}(|k_z(-\bm{\kappa}^{e}_{N_e})|^{\pm\frac{1}{2}}\mathbb{I}_2, \hdots, |k_z(\bm{\kappa}^{e}_{N_e})|^{\pm\frac{1}{2}}\mathbb{I}_2).\
\end{align}
In the case that $\Delta \kappa$ is different for different $k$-space regions, we can instead incorporate its different values into the matrices $\bm{\eta}^{\pm\frac{1}{2}}$.

To end this section, we note that it is possible to convert between the discrete scattering and transfer matrices. It can be shown from Eqs. (\ref{s1}) and (\ref{m111}) that
\begin{align}
	&\mathbf{r} = -\bm{\delta}^{-1}\bm{\gamma}, &&\bm{\alpha} = \mathbf{t} - \mathbf{r}'(\mathbf{t'})^{-1}\mathbf{r},\label{s2m1}\\
	&\mathbf{r}' =\bm{\beta}\bm{\delta}^{-1}, &&\bm{\beta} = \mathbf{r}'(\mathbf{t}')^{-1},\\
	&\mathbf{t} = \bm{\alpha} - \bm{\beta}\bm{\delta}^{-1}\bm{\gamma},  &&\bm{\gamma} = -(\mathbf{t}')^{-1}\mathbf{r},\\	
	&\mathbf{t}' =\bm{\delta}^{-1},  &&\bm{\delta} = (\mathbf{t}')^{-1},\label{s2m2}\
\end{align}
where, for example,
\begin{align}
	\mathbf{r} = \begin{pmatrix}
		\mathbf{r}_{pp} &\mathbf{r}_{pe}\\
		\mathbf{r}_{ep} & \mathbf{r}_{ee}\
	\end{pmatrix}
\end{align}
and the other matrices are defined analogously. These equations also hold for the normalized scattering and transfer matrices. 
\subsection{Conservation of Energy}
We now aim to derive the constraints imposed upon $\bar{\mathbf{S}}$ and $\bar{\mathbf{M}}$ by energy conservation. We begin by discretizing the conservation of energy equation for the continuous scattering matrix, namely Eq. (\ref{eq:sunit}). As before, we assume that the continuous scattering matrix is constant over each $k$-space region. The Dirac delta function $\delta(\bm\kappa' - \bm\kappa'')$, whose integral is by definition unity, can be replaced by the normalized Kronecker delta $\delta_{ij}/\Delta\kappa$. Furthermore, by using the normalized scattering matrix, it is possible to remove all $k_z$ terms from the equation. Note that a factor of $i$ is introduced whenever a $k_z$ factor corresponds to an evanescent wave. Replacing the integral with a sum and rewriting $\bm\kappa, \bm\kappa'$ and $\bm\kappa''$ as $\bm\kappa_l, \bm\kappa_i$ and $\bm\kappa_j$ respectively, we obtain
\begin{align}\label{eq:sunit_disc}
	\begin{split}
		&\sum_{\bm\kappa_l \in K_p}\bar{\mathbf{S}}^{\dagger}(\bm{\kappa}_l, \bm{\kappa}_i)\bar{\mathbf{S}}(\bm{\kappa}_l, \bm{\kappa}_j) \\
		&\quad= \begin{cases}
			\delta_{ij}\mathbb{I}_6\, & \text{if $\bm{\kappa}_i \in K_p$,  $\bm{\kappa}_j \in K_p$}, \\
			i\bar{\mathbf{S}}^{\dagger}(\bm{\kappa}_j, \bm{\kappa}_i)\, & \text{if $\bm{\kappa}_i \in K_p$, $\bm{\kappa}_j \in K_e$}, \\
			-i\bar{\mathbf{S}}(\bm{\kappa}_i, \bm{\kappa}_j)\, & \text{if $\bm{\kappa}_i \in K_e$, $\bm{\kappa}_j \in K_p$}, \\
			i\bar{\mathbf{S}}^{\dagger}(\bm{\kappa}_j, \bm{\kappa}_i) - i\bar{\mathbf{S}}(\bm{\kappa}_i, \bm{\kappa}_j) & \text{if $\bm{\kappa}_i \in K_e$, $\bm{\kappa}_j \in K_e$}.
		\end{cases}
	\end{split}
\end{align}

We can consider each of the four cases in Eq. (\ref{eq:sunit_disc}) separately. In each case, we may form four matrix equations by considering different sub-matrix blocks individually. Since there are a lot of equations and the algebra involved is rather lengthy and repetitive, we shall only present a single example here. Suppose $\bm\kappa_i, \bm\kappa_j \in K_p$ and we consider the first case of Eq. (\ref{eq:sunit_disc}). Equating the top-left blocks of the matrices on either side, we obtain
\begin{align}\label{step1}
	\sum_{\bm{\kappa}_l \in K_p}\big[ \bar{\mathbf{r}}^\dagger(\bm{\kappa}_l,\bm{\kappa}_i)\bar{\mathbf{r}}(\bm{\kappa}_l,\bm{\kappa}_j) + \bar{\mathbf{t}}^\dagger(\bm{\kappa}_l,\bm{\kappa}_i)\bar{\mathbf{t}}(\bm{\kappa}_l,\bm{\kappa}_j)\big] = \delta_{ij}\mathbb{I}_3.
\end{align}
We may further extract four equations from Eq. (\ref{step1}) by pre-multiplying and post-multiplying both sides by different combinations of $\mathbf{e}_\theta$ and $\mathbf{e}_\phi$. We also make use of the fact that, since $\mathbf{e}_k(\bm\kappa, \pm k_z), \mathbf{e}_\theta(\bm\kappa, \pm k_z)$ and $\mathbf{e}_\phi(\bm\kappa, \pm k_z)$ form an orthonormal basis of $\mathbb{C}^3$, we have \cite{lindell2015multiforms}
\begin{align}\label{step2}
	\mathbf{e}_k\mathbf{e}^\mathrm{T}_k + \mathbf{e}_\theta \mathbf{e}^\mathrm{T}_\theta + \mathbf{e}_\phi \mathbf{e}^\mathrm{T}_\phi = \mathbb{I}_3.
\end{align}
This means that multiplying any matrix by the combination of vectors on the left hand side of Eq. (\ref{step2}) will leave the matrix unchanged. Inserting this combination of vectors in between the matrix products in Eq. (\ref{step1}), pre-multiply the equation by $\mathbf{e}^{\mathrm{T}}_\theta(\bm\kappa_i, k_{zi})$ and post-multiplying the equation by $\mathbf{e}_\theta(\bm\kappa_j, k_{zj})$ yields
\begin{widetext}
\begin{align}\label{big}
	\begin{split}
		 \sum_{\bm\kappa_l \in K_p}\Big[\mathbf{e}^{\mathrm{T}}_\theta(\bm\kappa_i, k_{zi})\bar{\mathbf{r}}^\dagger(\bm{\kappa}_l,\bm{\kappa}_i)\big[\mathbf{e}_\theta(\bm\kappa_l, -k_{zl}) \mathbf{e}^\mathrm{T}_\theta(\bm\kappa_l, -k_{zl}) + \mathbf{e}_\phi(\bm\kappa_l, -k_{zl}) \mathbf{e}^\mathrm{T}_\phi(\bm\kappa_l, -k_{zl}) \big]\bar{\mathbf{r}}(\bm{\kappa}_l,\bm{\kappa}_j)\mathbf{e}_\theta(\bm\kappa_j, k_{zj})&\\
		 + \mathbf{e}^{\mathrm{T}}_\theta(\bm\kappa_i, k_{zi})\bar{\mathbf{t}}^\dagger(\bm{\kappa}_l,\bm{\kappa}_i)\big[\mathbf{e}_\theta(\bm\kappa_l, k_{zl}) \mathbf{e}^\mathrm{T}_\theta(\bm\kappa_l, k_{zl}) + \mathbf{e}_\phi(\bm\kappa_l, k_{zl}) \mathbf{e}^\mathrm{T}_\phi(\bm\kappa_l, k_{zl}) \big]\bar{\mathbf{t}}(\bm{\kappa}_l,\bm{\kappa}_j)\mathbf{e}_\theta(\bm\kappa_j, k_{zj})\Big]& = \delta_{ij}.		
	\end{split}
\end{align}
\end{widetext}
Note that all terms involving either $\mathbf{e}_k$ or $\mathbf{e}^{\mathrm{T}}_k$ vanish, as previously discussed. The left hand side of Eq. (\ref{big}) consists of eight bilinear forms, each of which can be simplified separately. For example, the first part of the first term in the sum in Eq. (\ref{big}) can be shown to be
\begin{align}
	\mathbf{e}^{\mathrm{T}}_\theta(\bm\kappa_i, k_{zi})\bar{\mathbf{r}}^\dagger(\bm{\kappa}_l,\bm{\kappa}_i)\mathbf{e}_\theta(\bm\kappa_l, -k_{zl}) = \bar{r}^*_{(l,i)\theta\theta},
\end{align}
where we use the fact that $\mathbf{e}_\theta = \mathbf{e}^*_\theta$ for propagating waves. Simplifying all terms in Eq. (\ref{big}) in a similar way gives
\begin{align}
	\begin{split}
\sum_{\bm\kappa_l \in K_p}\Big[&\bar{r}^*_{(l,i)\theta\theta}\bar{r}_{(l,j)\theta\theta}  + \bar{r}^*_{(l,i)\phi\theta}\bar{r}_{(l,j)\phi\theta} \\
+ &\bar{t}^*_{(l,i)\theta\theta}\bar{t}_{(l,j)\theta\theta}  + \bar{t}^*_{(l,i)\phi\theta}\bar{t}_{(l,j)\phi\theta}\Big] = \delta_{ij}
	\end{split}
\end{align} 
Repeating this process for all combinations of $\mathbf{e}_\theta$ and $\mathbf{e}_\phi$ and all sub-blocks in Eq. (\ref{eq:sunit_disc}) yields a large system of equations that can be shown to be equivalent to the single matrix equation
\begin{align}\label{sunit1}
	\bar{\mathbf{S}}^\dagger_{pp}\bar{\mathbf{S}}_{pp} = \mathbb{I}_{4N_p+2},
\end{align}
which is the classic result that a scattering matrix for a system that conserves energy and only considers propagating modes is unitary.

The other three cases for $\bm{\kappa}_i$ and $\bm{\kappa}_j$ can be treated similarly. After a lot of algebra, we obtain the equations
\begin{align}
		&\bar{\mathbf{S}}^{\dagger}_{pp}\bar{\mathbf{S}}_{pe} = i\bar{\mathbf{S}}^\dagger_{ep},\\
		&\bar{\mathbf{S}}^{\dagger}_{pe}\bar{\mathbf{S}}_{pe} = i(\bar{\mathbf{S}}^{\dagger}_{ee} - \bar{\mathbf{S}}_{ee}).\label{sunit2}
\end{align}
By introducing the matrices $\mathbb{I}_{p} = \mathrm{diag}(\mathbb{I}_{4N_p + 2}, \mathbb{O}_{4N_e})$ and $\mathbb{I}_{e} = \mathrm{diag}(\mathbb{O}_{4N_p + 2}, \mathbb{I}_{4N_e})$, we can combine Eqs. (\ref{sunit1})-(\ref{sunit2}) into the single equation
\begin{align}\label{SCOE}
	\bar{\mathbf{S}}^\dagger \mathbb{I}_{p}\bar{\mathbf{S}} = \mathbb{I}_{p} + i(\bar{\mathbf{S}}^\dagger \mathbb{I}_{e} - \mathbb{I}_{e}\bar{\mathbf{S}}),
\end{align}
which is the most general form of the conservation of energy constraint for the scattering matrix, incorporating both vector properties of the electric field as well as evanescent components. We note that our result here is consistent with a previously reported equation \cite{PhysRevE.72.026602}.

To derive the conservation of energy constraint for the discrete transfer matrix, we begin by discretizing Eq. (\ref{eq:munit}) to obtain
\begin{align}
	\begin{split}
		&\sum_{\bm{\kappa}_l \in K_p}\bar{\mathbf{M}}^{\dagger}(\bm{\kappa}_l, \bm{\kappa}_i)\bm{\Sigma}^z_{3}\bar{\mathbf{M}}(\bm{\kappa}_l, \bm{\kappa}_j) \\
		&\quad+  \sum_{\bm{\kappa}_l \in K_e}\bar{\mathbf{M}}^{\dagger}(\bm{\kappa}_l, \bm{\kappa}_i)\bm{\Sigma}^y_{3}\bar{\mathbf{M}}(\bm{\kappa}_l, \bm{\kappa}_j) \\
		&\quad\quad\quad= \begin{cases}
			\delta_{ij}\bm{\Sigma}^z_{3} & \text{if $\bm{\kappa}_i \in K_p$,  $\bm{\kappa}_j \in K_p$}, \\
			\delta_{ij}\bm{\Sigma}^y_{3} & \text{if $\bm{\kappa}_i \in K_e$,  $\bm{\kappa}_j \in K_e$}, \\
			\mathbb{O}_6 & \text{otherwise.}
		\end{cases}
	\end{split}
\end{align}
Performing similar steps to those demonstrated for the scattering matrix, we find, again after a lot of algebra, the equations
\begin{align}
		&\bar{\mathbf{M}}^{\dagger}_{pp}\mathbf{\Sigma}^z_{4N_p + 2}\bar{\mathbf{M}}_{pp} + \bar{\mathbf{M}}^{\dagger}_{ep}\mathbf{\Sigma}^y_{4N_e}\bar{\mathbf{M}}_{ep} = \mathbf{\Sigma}^z_{4N_p + 2},\label{m1}\\		&\bar{\mathbf{M}}^{\dagger}_{pp}\mathbf{\Sigma}^z_{4N_p + 2}\bar{\mathbf{M}}_{pe} + \bar{\mathbf{M}}^{\dagger}_{ep}\mathbf{\Sigma}^y_{4N_e}\bar{\mathbf{M}}_{ee} = \mathbb{O},\label{0mat}\\
		&\bar{\mathbf{M}}^{\dagger}_{pe}\mathbf{\Sigma}^z_{4N_p + 2}\bar{\mathbf{M}}_{pe} + \bar{\mathbf{M}}^{\dagger}_{ee}\mathbf{\Sigma}^y_{4N_e}\bar{\mathbf{M}}_{ee} = \mathbf{\Sigma}^y_{4N_e},\label{m2}
\end{align}
where the zero matrix in Eq. (\ref{0mat}) is of size $(4N_p + 2) \times 4N_e$. Eqs. (\ref{m1})-(\ref{m2}) can be combined into the single equation
\begin{align}\label{MCOE}
	\bar{\mathbf{M}}^{\dagger} \mathbf{\Omega}\bar{\mathbf{M}} = \mathbf{\Omega},
\end{align}
where $\mathbf{\Omega} = \mathrm{diag}(\mathbf{\Sigma}^z_{4N_p + 2},\mathbf{\Sigma}^y_{4N_e})$. Eq. (\ref{MCOE}) therefore represents conservation of energy for the discrete transfer matrix.
\subsection{Reciprocity/Time Reversal Symmetry}
As noted previously, when conservation of energy holds, reciprocity and time reversal symmetry are equivalent. In this section we shall therefore henceforth refer to both conditions as `reciprocity'. We begin by deriving the reciprocity constraint for the discrete scattering matrix. Note that for the continuous scattering matrix the reciprocity relation Eq. (\ref{eq:srec}) is considerably simpler than the time reversal symmetry relation Eq. (\ref{eq:strs}), particularly as it does not involve an integral. We therefore begin our derivation with Eq. (\ref{eq:srec}).

In terms of the normalized scattering matrix, the reciprocity constraint of Eq. (\ref{eq:srec}) has two different cases: $\bar{\mathbf{S}}(\bm{\kappa}_i, \bm{\kappa}_j) = \bar{\mathbf{S}}^\mathrm{T}(-\bm{\kappa}_j, -\bm{\kappa}_i)$ if either $\bm{\kappa}_i, \bm{\kappa}_j \in K_p$ or $\bm{\kappa}_i, \bm{\kappa}_j \in K_e$, and $\bar{\mathbf{S}}(\bm{\kappa}_i, \bm{\kappa}_j) = i\bar{\mathbf{S}}^\mathrm{T}(-\bm{\kappa}_j, -\bm{\kappa}_i)$ otherwise. Let us consider first the case where $\bm{\kappa}_i, \bm{\kappa}_j \in K_p$. As in the previous section, we again consider each sub-matrix of $\bar{\mathbf{S}}$ separately. Comparing top-left blocks, we have
\begin{align}\label{int1}
	\bar{\mathbf{r}}(\bm{\kappa}_i, \bm{\kappa}_j) = \bar{\mathbf{r}}(-\bm{\kappa}_j, -\bm{\kappa}_i).
\end{align}
If we pre-multiply Eq. (\ref{int1}) by $\mathbf{e}^{\mathrm{T}}_{\theta}(\bm{\kappa}_i, -k_{zi})$ and post-multiply by $\mathbf{e}_{\theta}(\bm{\kappa}_j, k_{zj})$, the left hand side becomes $r_{(i,j)\theta \theta}$ and the right hand side, after a bit of manipulation, becomes $r_{(-j,-i)\theta \theta}$ where here the subscript $-j$ refers to the mode with transverse wavevector $-\bm{\kappa}_j$. Repeating this for all four combinations of $\mathbf{e}_\theta$ and $\mathbf{e}_\phi$, we obtain the relations
\begin{align}
		&\bar{r}_{(i,j)\theta \theta} = \bar{r}_{(-j,-i)\theta \theta},\label{rrec1}, \quad &&\bar{r}_{(i,j)\theta \phi} = -\bar{r}_{(-j,-i)\phi \theta},\\
		&\bar{r}_{(i,j)\phi \theta} = -\bar{r}_{(-j,-i)\theta \phi}, \quad &&\bar{r}_{(i,j)\phi \phi} = \bar{r}_{(-j,-i)\phi \phi}.\label{rrec4}\
\end{align} 
Eqs. (\ref{rrec1}) and (\ref{rrec4}) are equivalent to the single matrix relation
\begin{align}\label{rrsym}
	\bar{\mathbf{r}}_{(i,j)} = \bar{\mathbf{r}}^\textrm{R}_{(-j,-i)},
\end{align}
where we have introduced the reciprocal operator R, which we define such that if $[\mathbf{A}]_{mn}$ is the $(m,n)$ element of the matrix $\mathbf{A}$, then
\begin{align}\label{rdef}
	[\mathbf{A}^\textrm{R}]_{mn} = [\mathbf{A}]_{nm}(-1)^{m+n}.
\end{align} 
This particular symmetry of the reflection matrix is a well-known result in scattering theory for polarized light and is sometimes referred to as the backscattering theorem \cite{van2012light}. The operator $\mathrm{R}$ has also been discussed previously in the context of reciprocal Jones matrices \cite{Bhandari:08}. 

By now carefully considering the structure of the matrix $\bar{\mathbf{r}}_{pp}$, it follows from  Eq. (\ref{rrsym}) that
\begin{align}\label{rppeq}
	\begin{split}
		\bar{\mathbf{r}}_{pp} = \bm{\sigma}_p \bar{\mathbf{r}}^{\textrm{R}}_{pp} \bm{\sigma}_p,
	\end{split}
\end{align}
where 
\begin{align}
	\bm{\sigma}_p = \begin{pmatrix}
		 \mathbb{O}& &\mathbb{I}_2\\
		 &\reflectbox{$\ddots$} & \\
		\mathbb{I}_2 & &\mathbb{O}\
	\end{pmatrix}
\end{align}
is the matrix containing $2N_p + 1$ copies of the $2\times2$ identity matrix on its anti-diagonal and zeroes elsewhere. The effect of multiplying a matrix on either side by $\bm{\sigma}_p$ is to reflect the positions of all $2\times 2$ sub-matrices horizontally and vertically about the central rows and columns of the matrix, but to leave the sub-matrices themselves unchanged. This is necessary so that Eq. (\ref{rppeq}) correctly equates sub-matrices of $\mathbf{r}_{pp}$ that are related by an inversion of transverse wavevectors. Similarly, by considering the other sub-matrices of $\bar{\mathbf{S}}$, we find
\begin{align}
		&\bar{\mathbf{t}}_{pp} = \bm{\sigma}_p \bar{\mathbf{t}}'^{\textrm{R}}_{pp}\bm{\sigma}_p,\\
		&\bar{\mathbf{r}}'_{pp} = \bm{\sigma}_p \bar{\mathbf{r}}'^{\textrm{R}}_{pp}\bm{\sigma}_p,
\end{align}
which can be combined into the single equation
\begin{align}\label{sr1}
	\bar{\mathbf{S}}_{pp} = \bm{\sigma}_{pp}\bar{\mathbf{S}}^{\textrm{R}}_{pp}\bm{\sigma}_{pp},
\end{align}
where $\bm{\sigma}_{pp} = \textrm{diag}(\bm{\sigma}_p, \bm{\sigma}_p)$. If we similarly introduce $\bm{\sigma}_e$ as the matrix containing $2N_e$ copies of $\mathbb{I}_2$ on its anti-diagonal and zeroes elsewhere, and $\bm{\sigma}_{ee} = \textrm{diag}(\bm{\sigma}_e, \bm{\sigma}_e)$, we can further derive
\begin{align}
		&\bar{\mathbf{S}}_{pe} = \bm{\sigma}_{pp}i\bar{\mathbf{S}}^{\textrm{R}}_{ep}\bm{\sigma}_{ee},\\
			&\bar{\mathbf{S}}_{ee} = \bm{\sigma}_{ee}\bar{\mathbf{S}}^{\textrm{R}}_{ee}\bm{\sigma}_{ee}.\label{sr2}\
\end{align}
Finally, Eqs. (\ref{sr1})-(\ref{sr2}) can be combined into the single equation
\begin{align}\label{SREC}
	\bar{\mathbf{S}} = \bm{\omega}^*\bar{\mathbf{S}}^{\textrm{R}}\bm{\omega},
\end{align}
where $\bm{\omega} = \textrm{diag}(\bm{\sigma}_{pp}, i\bm{\sigma}_{ee})$.
Eq. (\ref{SREC}) is the reciprocity constraint for the scattering matrix, which generalizes the well known equation $\mathbf{S} = \mathbf{S}^{\mathrm{T}}$.

To derive the reciprocity constraint for the transfer matrix, we begin with the time reversal invariance constraint Eq. (\ref{trstrans}), which is notably simpler than the reciprocity constraint in Eq. (\ref{eq:transrec}), and perform analogous steps to those used in the derivation for the scattering matrix. We eventually arrive at the equations
\begin{align}
	&\bar{\mathbf{M}}_{pp} = \bm{\sigma}'_{pp}\bar{\mathbf{M}}^{\dagger \textrm{R}}_{pp} \bm{\sigma}'_{pp},\label{fin1} \quad &&\bar{\mathbf{M}}_{pe} = \bm{\sigma}'_{pp}\bar{\mathbf{M}}^{\dagger \textrm{R}}_{pe} \bm{\sigma}_{ee},\\
	&\bar{\mathbf{M}}_{ep} = \bm{\sigma}_{ee}\bar{\mathbf{M}}^{\dagger \textrm{R}}_{ep} \bm{\sigma}'_{pp}, \quad &&\bar{\mathbf{M}}_{ee} = \bm{\sigma}_{ee}\bar{\mathbf{M}}^{\dagger \textrm{R}}_{pp} \bm{\sigma}_{ee},\label{fin4}\
\end{align}
where
\begin{align}
	\bm{\sigma}'_{pp} = \bm{\Sigma}^{x}_{4N_p + 2} \bm{\sigma}_{pp}= \begin{pmatrix}
		\mathbb{O}&\bm{\sigma}_{p}\\
		\bm{\sigma}_{p}&\mathbb{O}\
	\end{pmatrix}.
\end{align}
Eqs. (\ref{fin1})-(\ref{fin4}) can be combined into the single equation
\begin{align}\label{MREC}
	\bar{\mathbf{M}} = \bm{\omega}'\bar{\mathbf{M}}^{\dagger \textrm{R}}\bm{\omega}', 
\end{align}
where $\bm{\omega}' = \textrm{diag}(\bm{\sigma}'_{pp}, \bm{\sigma}_{ee})$. Eq. (\ref{MREC}) is hence the reciprocity constraint for the discrete transfer matrix. 

Eqs. (\ref{SCOE}), (\ref{MCOE}), (\ref{SREC}) and (\ref{MREC}) constitute the main results of our work and must be satisfied for any system that conserves energy and is reciprocal/time reversal invariant. We note that all four of these equations are algebraic and are therefore much simpler to work with than the integral constraints for the continuous scattering and transfer matrices.

\subsection{Comparison With Previous Results}
Here we briefly compare our results with those that already exist in the literature. In particular, we show that the more commonly presented equations follow from our results as special cases. From Refs \cite{RevModPhys.89.015005}, \cite{MELLO1988290}, \cite{RevModPhys.69.731}, \cite{doi:10.1063/1.531112} and others we see that, for a system that conserves energy and and is reciprocal/time reversal invariant, the matrices $\bar{\mathbf{S}}$ and $\bar{\mathbf{M}}$ obey the equations
\begin{align}
	\bar{\mathbf{S}}\bar{\mathbf{S}}^{\dagger} &= \mathbb{I},\label{theirs1}\\
	\bar{\mathbf{S}} &= \bar{\mathbf{S}}^\mathrm{T},\label{theirs2}\\
	\bar{\mathbf{M}}^\dagger \mathbf{\Sigma}^z \bar{\mathbf{M}} &= \mathbf{\Sigma}^z,\label{theirs3}\\
	\mathbf{\Sigma}^x \bar{\mathbf{M}} \mathbf{\Sigma}^x\label{theirs4} &= \bar{\mathbf{M}}^*, 
\end{align}
where $\mathbb{I}, \mathbf{\Sigma}^x$ and $\mathbf{\Sigma}^z$ are of the appropriate size. 

There are three important differences between the matrices in Eqs. (\ref{theirs1})-(\ref{theirs4}) and those of our results. Firstly, the scattering and transfer matrices in Eqs. (\ref{theirs1})-(\ref{theirs4}) are defined only for propagating modes in the far field. Therefore, in comparing with our results, it is sufficient to only consider $\bar{\mathbf{S}}_{pp}$ and $\bar{\mathbf{M}}_{pp}$. As previously noted, Eq. (\ref{sunit1}) is identical to the unitarity condition of Eq. (\ref{theirs1}). For the transfer matrix, since in the far field we have $\bar{\mathbf{M}}_{ep} = \mathbb{O}$, we see that Eq. (\ref{m1}) reduces to the same form as of Eq. (\ref{theirs3}).

The second important difference is that Eqs. (\ref{theirs1})-(\ref{theirs4}) are defined only for scalar waves. A scalar wave formalism is appropriate when there is no change in polarisation state induced by the scattering medium. For this to be the case within our formalism, it is necessary for each $2\times 2$ sub-matrix within $\bar{\mathbf{S}}$ and $\bar{\mathbf{M}}$ to have zero off-diagonal elements. Mathematically, this means that $[\bar{\mathbf{S}}]_{mn} = [\bar{\mathbf{M}}]_{mn} = 0$ when $m+n$ is an odd number. By now inspecting Eq. (\ref{rdef}) we see that, if this is the case, then the reciprocal operator is identical to a regular matrix transpose and in all of our previous equations we may make the notational transformation $\mathrm{R} \to \mathrm{T}$.

The final difference is that Eqs. (\ref{theirs1})-(\ref{theirs4}) are defined in a quasi-one-dimensional geometry for which each mode has a wavevector with a unique $z$ component. Consequently, there are no two modes whose wavevectors have the same $z$ components, but different transverse components. More specifically, this means that there is no distinction between the two modes with wavevectors $(\bm{\kappa}, k_z)^\mathrm{T}$ and $(-\bm{\kappa}, k_z)^\mathrm{T}$. If we were to enforce this constraint within our formalism, all transverse wavevectors in the sets defined in Eqs. (\ref{modes1}) and (\ref{modes2}) containing a minus sign would become extraneous and could be removed. Moreover, the use of the matrix $\bm{\sigma}_p$ to correctly associate sub-matrices of $\bar{\mathbf{S}}$ and $\bar{\mathbf{M}}$ that describe scattering between modes with inverse transverse wavevectors would no longer be necessary and $\bm{\sigma}_p$ would be replaced by the identity matrix of the appropriate size. Making this change, combined with the previous change regarding the reciprocal operator, transforms Eq. (\ref{sr1}) into Eq. (\ref{theirs2}). Finally, we see that under these changes we also have $\bm{\sigma}'_{pp} \to \bm{\Sigma}_x$, $\bar{\mathbf{M}}^{\dagger \mathrm{R}} \to \bar{\mathbf{M}}^*$ and hence Eq. (\ref{fin1}) becomes identical to Eq. (\ref{theirs4}), completing the comparison.

\subsection{Numerical Example: Glass-Air Interface}
In this section we demonstrate the validity of our results with a numerical example. We consider a planar glass-air boundary with glass on the left and air on the right. We choose $n_g = 1.5$ and $n_a = 1.0$ to be the refractive indices of the glass and air respectively. In the glass, we consider three right-travelling modes with wavevectors $\mathbf{k}_1$, $\mathbf{k}_2$ and $\mathbf{k}_3$ whose $z$ components are $k_{z1}/k_0 = 1.41$, $k_{z2}/k_0 = 0.90$ and $k_{z3}/k_0 = 0.73i$, where each numerical value is given to two decimal places. In the air, we consider the set of refracted wavevectors $\mathbf{k}'_1$, $\mathbf{k}'_2$ and $\mathbf{k}'_3$, whose $z$ components $k'_{z1}/k_0 = 0.87$, $k'_{z2}/k_0 = 0.66i$ and $k'_{z3}/k_0 = 1.33i$ follow from those of $\mathbf{k}_1, \mathbf{k}_2$ and $\mathbf{k}_3$ by Snell's law. We identify the `scattering medium' in this example with the planar boundary and note that our choice of wavevectors incorporates different types of scattering. For example, the mode with wavevector $\mathbf{k}_1$ is partially transmitted and reflected, but the mode with wavevector $\mathbf{k}_2$ is totally internally reflected. We describe left-travelling modes in the glass medium using the reflected wavevectors $\widetilde{\mathbf{k}}_1$, $\widetilde{\mathbf{k}}_2$ and $\widetilde{\mathbf{k}}_3$ and, similarly, we describe left-travelling modes in the air using the wavevectors $\widetilde{\mathbf{k}}'_1$, $\widetilde{\mathbf{k}}'_2$ and $\widetilde{\mathbf{k}}'_3$. For each wavevector mode, we consider both $s$ and $p$ polarisations (we shall use $s$ and $p$ to refer to polarization states in this section alone) with the usual definitions for a planar interface and write, $\mathbf{k}_{1s}$ to refer to an $s$ polarized wave with wavevector $\mathbf{k}_1$ and similarly for $\mathbf{k}_{1p}$. 

We calculate the scattering matrix element-wise using the standard Fresnel equations (see e.g. Ref. \cite{born1999principles}). We then normalize each element of the scattering matrix according to Eq. (\ref{norm1}), which gives
\begin{widetext}
\begin{align}
   \bar{\mathbf{S}} = \; \begin{blockarray}{ccccccccccccc@{\hskip 0.4in}}
   	 \mathbf{k}_{1s} & \mathbf{k}_{1p} & \mathbf{k}_{2s} & \mathbf{k}_{2p} & \widetilde{\mathbf{k}}'_{1s} & \widetilde{\mathbf{k}}'_{1p} & \mathbf{k}_{3s} & \mathbf{k}_{3p} & \widetilde{\mathbf{k}}'_{2s} & \widetilde{\mathbf{k}}'_{2p} & \widetilde{\mathbf{k}}'_{3s} & \widetilde{\mathbf{k}}'_{3p} &  \\
   	\begin{block}{(cccc|cc|cc|cccc)c}
			0.24 & 0 & 0 & 0 & 0.97 & 0 & 0 & 0 & 0 & 0 & 0 & 0 & \widetilde{\mathbf{k}}_{1s}\\
			0 & -0.16 & 0 & 0 & 0 & 0.99 & 0 & 0 & 0 & 0 & 0 & 0 & \widetilde{\mathbf{k}}_{1p}\\
			0 & 0 & 0.30-0.96i & 0 & 0 & 0 & 0 & 0 & 0.82+1.11i & 0 & 0 & 0 & \widetilde{\mathbf{k}}_{2s} \\
			0 & 0 & 0 & -0.47-0.88i & 0 & 0 & 0 & 0 & 0 & 1.14+0.69i & 0 & 0 & \widetilde{\mathbf{k}}_{2p}\\
			\cline{1-12}
			0.97 & 0 & 0 & 0 & -0.24 & 0 & 0 & 0 & 0 & 0 & 0 & 0 & \mathbf{k}'_{1s} \\
			0 & 0.99 & 0 & 0 & 0 & 0.16 & 0 & 0 & 0 & 0 & 0 & 0 & \mathbf{k}'_{1p} \\
			\cline{1-12}
			0 & 0 & 0 & 0 & 0 & 0 & -0.29 & 0 & 0 & 0 & 0.96 & 0 & \widetilde{\mathbf{k}}_{3s} \\
			0 & 0 & 0 & 0 & 0 & 0 & 0 & -0.61 & 0 & 0 & 0 & 0.79 & \widetilde{\mathbf{k}}_{3p} \\
			\cline{1-12}
			0 & 0 & 1.11-0.82i & 0 & 0 & 0 & 0 & 0 & -0.30+0.96i & 0 & 0 & 0 &\mathbf{k}'_{2s} \\
			0 & 0 & 0 & 0.69-1.14i & 0 & 0 & 0 & 0 & 0 & 0.47+0.88i & 0 & 0 &\mathbf{k}'_{2p}\\
			0 & 0 & 0 & 0 & 0 & 0 & 0.96 & 0 & 0 & 0 & 0.29 & 0 & \mathbf{k}'_{3s}\\
			0 & 0 & 0 & 0 & 0 & 0 & 0 & 0.79 & 0 & 0 & 0 & 0.61 & \mathbf{k}'_{3p}\\
	\end{block}
\end{blockarray},
\end{align}
\end{widetext}
where we have indicated the incident and outgoing modes above and to the right of the matrix respectively. We have included horizontal and vertical lines within the matrix as a visual aid to map out the block structure of $\bar{\mathbf{S}}$ as in Eq. (\ref{sblock}). We verify numerically that $\bar{\mathbf{S}}$ satisfies the conservation of energy constraint given by Eq. (\ref{SCOE}) with $\mathbb{I}_{p} = \mathrm{diag}(\mathbb{I}_6, \mathbb{O}_6)$ and $\mathbb{I}_{e} = \mathrm{diag}(\mathbb{O}_6, \mathbb{I}_6)$. 

In this particular example, the scattering matrix includes evanescent components, but contains no polarization mixing and does not consider pairs of inverse transverse wavevector modes. As per the discussion in the previous section, the reciprocity constraint for the scattering matrix can therefore be simplified. Specifically, in Eq. (\ref{SREC}) we may make the transformations $\mathrm{R} \to \mathrm{T}$ and $\bm{\sigma}_{pp}, \bm{\sigma}_{ee} \to \mathbb{I}_6$, which yields the modified equation 
\begin{align}\label{smodified}
	\bar{\mathbf{S}} = \bm{\omega}^* \bar{\mathbf{S}}^\mathrm{T}\bm{\omega},
\end{align}
where, in this instance, $\bm{\omega} = \mathrm{diag}(\mathbb{I}_6, i\mathbb{I}_6)$. Eq. (\ref{smodified}) is also satisfied by our scattering matrix $\bar{\mathbf{S}}$.

We calculate the transfer matrix from the scattering matrix using Eqs. (\ref{s2m1})-(\ref{s2m2}). This gives
\begin{widetext}
\begin{align}
   \bar{\mathbf{M}} = \;\begin{blockarray}{ccccccccccccc@{\hskip 0.4in}}
	\mathbf{k}_{1s} & \mathbf{k}_{1p} & \mathbf{k}_{2s} & \mathbf{k}_{2p} & \widetilde{\mathbf{k}}_{1s} & \widetilde{\mathbf{k}}_{1p} & \widetilde{\mathbf{k}}_{2s} &\widetilde{\mathbf{k}}_{2p} & \mathbf{k}_{3s} & \mathbf{k}_{3p} & \widetilde{\mathbf{k}}_{3s} & \widetilde{\mathbf{k}}_{3p} &  \\
		\begin{block}{(cccc|cccc|cc|cc)c}
		1.03 & 0 & 0 & 0 & -0.25 & 0 & 0 & 0 & 0 & 0 & 0 & 0 & \mathbf{k}'_{1s}\\
		0 & 1.01 & 0 & 0 & 0 & 0.16 & 0 & 0 & 0 & 0 & 0 & 0 & \mathbf{k}'_{1p}\\
		\cline{1-12}
		-0.25 & 0 & 0 & 0 & 1.03 & 0 & 0 & 0 & 0 & 0 & 0 & 0 & \widetilde{\mathbf{k}}'_{1s} \\
		0 & 0.16 & 0 & 0 & 0 & 1.01 & 0 & 0 & 0 &0 & 0 & 0 & \widetilde{\mathbf{k}}'_{1p}\\
		\cline{1-12}
		0 & 0 & 0.43 - 0.58i & 0 & 0 & 0 & 0.43+0.58i & 0 & 0 & 0 & 0 & 0 & \mathbf{k}'_{2s} \\
		0 & 0 & 0 & 0.64-0.39i & 0 & 0 & 0 & 0.64+0.39i & 0 & 0 & 0 & 0 & \mathbf{k}'_{2p} \\
		0 & 0 & 0 & 0 & 0 & 0 & 0 & 0 & 1.05 & 0 & 0.31 & 0 & \mathbf{k}'_{3s} \\
		0 & 0 & 0 & 0 & 0 & 0 & 0 & 0 & 0 & 1.26 & 0 & 0.77 &  \mathbf{k}'_{3p} \\
		\cline{1-12}
		0 & 0 & 0.43 + 0.58i & 0 & 0 & 0 & 0.43-0.58i & 0 & 0 & 0 & 0 & 0 & \widetilde{\mathbf{k}}'_{2s} \\
		0 & 0 & 0 & 0.64 + 0.39i & 0 & 0 & 0 & 0.64 - 0.39i & 0 & 0 & 0 & 0 & \widetilde{\mathbf{k}}'_{2p}\\
		0 & 0 & 0 & 0 & 0 & 0 & 0 & 0 & 0.31 & 0 & 1.05 & 0 & \widetilde{\mathbf{k}}'_{3s}\\
		0 & 0 & 0 & 0 & 0 & 0 & 0 & 0 & 0 & 0.77 & 0 & 1.26 & \widetilde{\mathbf{k}}'_{3p}\\
	\end{block}
\end{blockarray},\
\end{align}
\end{widetext}
where, as before, the horizontal and vertical lines show the block structure of $\bar{\mathbf{M}}$ as in Eq. (\ref{mblock}). Since there is an uneven number of propagating and evanescent modes on either side of the boundary, $\bar{\mathbf{M}}$ has an irregular block structure and, notably, the matrices $\bar{\mathbf{M}}_{pp}$ and $\bar{\mathbf{M}}_{ee}$ are no longer square. It is therefore necessary to modify the sizes of the sub-matrices of $\bm{\Omega}$ and $\bm{\omega}'$ in Eqs. (\ref{MCOE}) and (\ref{MREC}) to account for this asymmetry. Furthermore, in addition to the transformations made for the scattering matrix reciprocity equation, we also replace $\bm{\sigma}'_{pp} \to \bm{\Sigma}^x$. Ultimately, we find that $\bar{\mathbf{M}}$ must satisfy
\begin{align}
	\bar{\mathbf{M}^{\dagger}} \mathbf{\Omega}_1\bar{\mathbf{M}} &= \mathbf{\Omega}_2,\label{last1}\\
	\bm{\omega}'_1\bar{\mathbf{M}}^{*}\bm{\omega}'_2 &= \bar{\mathbf{M}}  , \label{last2}
\end{align}
where 
\begin{align}
	&\bm{\Omega}_1 = \begin{pmatrix}
		\bm{\Sigma}^z_2 & \mathbb{O}\\
		\mathbb{O} & \bm{\Sigma}^y_4\
		\end{pmatrix}, \quad && \bm{\Omega}_2 = \begin{pmatrix}
		\bm{\Sigma}^z_4 & \mathbb{O}\\
		\mathbb{O} & \bm{\Sigma}^y_2\
	\end{pmatrix},\\
	&\bm{\omega}'_1 = \begin{pmatrix}
	\bm{\Sigma}^x_2 & \mathbb{O}\\
	\mathbb{O} & \mathbb{I}_8\
	\end{pmatrix}, \quad && \bm{\omega}'_2 = \begin{pmatrix}
	\bm{\Sigma}^x_4 & \mathbb{O}\\
	\mathbb{O} & \mathbb{I}_4\
	\end{pmatrix}.
\end{align}
Despite these modifications, we note that Eqs. (\ref{last1}) and (\ref{last2}) still have the same basic form as Eqs. (\ref{MCOE}) and (\ref{MREC}). Numerical calculations confirm that Eqs. (\ref{last1}) and (\ref{last2}) are indeed satisfied.

\section{Conclusion}
In this paper we have derived the general set of constraints for the scattering and transfer matrices imposed by conservation of energy, reciprocity and time reversal symmetry. Our formalism considers the general case of vectorial light and allows for fields containing evanescent components. We have extended previously known results, such as in Eqs. (\ref{theirs1})-(\ref{theirs4}), and have introduced a more general set of constraints for the discrete scattering and transfer matrices, which are given in Eqs. (\ref{SCOE}), (\ref{MCOE}), (\ref{SREC}) and (\ref{MREC}). We have discussed the differences between our results and Eqs. (\ref{theirs1})-(\ref{theirs4}) and have demonstrated that the latter can be regarded as special cases of the former. In particular, Eqs. (\ref{theirs1})-(\ref{theirs4}) follow from our results when one is able to neglect evanescent field components, polarization mixing and a distinction between inverse transverse wavevector modes. We have given a demonstration of the validity of our results in the simple example of the scattering of polarized light at a planar glass-air interface, including the effect of total internal reflection. In our formalism, we have made minimal assumptions regarding the nature of the scattering medium and our results should therefore be valid for a wide range of systems.

Given the popularity of matrix methods in scattering studies and the increasing ease with which it is possible to experimentally determine scattering and transfer matrices, we believe that our results will serve as useful guides for future experimental research. We also believe that our results will help define the limits for physically realizable scattering and transfer matrices in matrix-based numerical simulations and theoretical studies of optical scattering. Our results should be particularly important when one is interested in a description of a scattering problem that incorporates any combination of evanescent modes, vectorial light and reciprocity/time reversal symmetry in a system of more than one dimension. 

\section*{Acknowledgements}
This work was funded by the Royal Society.

\appendix
\section{Reduced Matrices}
To complement Eq. (\ref{app}) in the main text, we present here a list of equations for the $\theta$ and $\phi$ components of the reduced versions of each of the matrices $\mathbf{r}$, $\mathbf{r}', \mathbf{t}, \mathbf{t}', \bm{\alpha},\bm{\beta}, \bm{\gamma}$ and $\bm{\delta}$. In each equation, the subscripts $m$ and $n$ represent either $k$, $\theta$ or $\phi$. We have
\begin{align}
	&	r_{(i,j)mn} = \mathbf{e}^{\mathrm{T}}_{m}(\bm{\kappa}_i, -k_{zi}) \mathbf{r}(\bm{\kappa}_i, \bm{\kappa}_j) \mathbf{e}_{n}(\bm{\kappa}_j, k_{zj}),\\
		&r'_{(i,j)mn} = \mathbf{e}^{\mathrm{T}}_{m}(\bm{\kappa}_i, k_{zi}) \mathbf{r}'(\bm{\kappa}_i, \bm{\kappa}_j) \mathbf{e}_{n}(\bm{\kappa}_j, -k_{zj}),\\
		&t_{(i,j)mn} = \mathbf{e}^{\mathrm{T}}_{m}(\bm{\kappa}_i, 	k_{zi}) \mathbf{t}(\bm{\kappa}_i, \bm{\kappa}_j) \mathbf{e}_{n}(\bm{\kappa}_j, k_{zj}),\\
		&t'_{(i,j)mn} = \mathbf{e}^{\mathrm{T}}_{m}(\bm{\kappa}_i, -k_{zi}) \mathbf{t}'(\bm{\kappa}_i, \bm{\kappa}_j) \mathbf{e}_{n}(\bm{\kappa}_j, -k_{zj}),\\
		&\alpha_{(i,j)mn} = \mathbf{e}^{\mathrm{T}}_{m}(\bm{\kappa}_i, k_{zi}) \bm{\alpha}(\bm{\kappa}_i, \bm{\kappa}_j) \mathbf{e}_{n}(\bm{\kappa}_j, k_{zj}),\\
		&\beta_{(i,j)mn} = \mathbf{e}^{\mathrm{T}}_{m}(\bm{\kappa}_i, k_{zi}) \bm{\beta}(\bm{\kappa}_i, \bm{\kappa}_j) \mathbf{e}_{n}(\bm{\kappa}_j, -k_{zj}),\\
		&\gamma_{(i,j)mn} = \mathbf{e}^{\mathrm{T}}_{m}(\bm{\kappa}_i, -k_{zi}) \bm{\gamma}(\bm{\kappa}_i, \bm{\kappa}_j) \mathbf{e}_{n}(\bm{\kappa}_j, k_{zj}),\\
		&\delta_{(i,j)mn} = \mathbf{e}^{\mathrm{T}}_{m}(\bm{\kappa}_i, -k_{zi}) \bm{\delta}(\bm{\kappa}_i, \bm{\kappa}_j) \mathbf{e}_{n}(\bm{\kappa}_j, -k_{zj}).\
\end{align}
As discussed in the main text, if $\mathbf{A}(\bm{\kappa}_i, \bm{\kappa}_j)$ refers to any of the above eight matrices, then 
\begin{align}
	\mathbf{e}^{\mathrm{T}}_k(\bm{\kappa}_i, \pm k_{zi}) \mathbf{A}(\bm{\kappa}_i, \bm{\kappa}_j) &= \mathbf{0}^{\mathrm{T}},\\
	\mathbf{A}(\bm{\kappa}_i, \bm{\kappa}_j)\mathbf{e}_k(\bm{\kappa}_j, \pm k_{zj}) &= \bm{0},
\end{align}
where the plus or minus sign is chosen according to the choice of $\mathbf{A}$. 

\providecommand{\noopsort}[1]{}\providecommand{\singleletter}[1]{#1}%

\end{document}